\documentclass[1p, preprint]{elsarticle}

\usepackage{hyperref}
\journal{Journal of \LaTeX\ Templates}

\biboptions{authoryear}
\usepackage{graphicx}
\usepackage{xcolor}
\usepackage{amsmath}
\usepackage{amssymb}
\usepackage{url}
\usepackage{multirow}
\usepackage{xcolor}
\usepackage{makecell}
\usepackage{booktabs}
\usepackage{tabu}

\newcommand\myeq{\mathrel{\overset{\makebox[0pt]{\mbox{\normalfont\tiny\sffamily def}}}{=}}}

\begin{document}

\begin{frontmatter}

\title{Lung Swapping Autoencoder: Learning a Disentangled Structure-texture Representation of Chest Radiographs}
% \tnotetext[mytitlenote]{Fully documented templates are available in the elsarticle package on \href{http://www.ctan.org/tex-archive/macros/latex/contrib/elsarticle}{CTAN}.}

% \author{Authors\fnref{myfootnote}}
% \address{Radarweg 29, Amsterdam}
% \fntext[myfootnote]{Since 1880.}

\author[1]{Lei Zhou\corref{mycorrespondingauthor}}
\ead{lezzhou@cs.stonybrook.edu}
\author[2]{Joseph Bae}
\author[1]{Huidong Liu}
\author[3]{Gagandeep Singh}
\author[3]{Jeremy Green}
\author[4]{\\Amit Gupta}
\author[1]{Dimitris Samaras}
\author[2]{Prateek Prasanna}
\address[1]{Department of Computer Science, Stony Brook University, NY, USA}
\address[2]{Department of Biomedical Informatics, Stony Brook University, NY, USA}
\address[3]{Department of Radiology, Newark Beth Israel Medical Center, NJ, USA}
\address[4]{Department of Radiology, University Hospitals Cleveland Medical Center, OH, USA}
% \affil[2]{Department of Biomedical Informatics, Stony Brook University, NY, USA}
% \author[Department of Computer Science, Stony Brook University, NY, USA]{Lei Zhou\corref{mycorrespondingauthor}}
% \author[Department of Biomedical Informatics, Stony Brook University, NY, USA]{Joseph Bae}
% \author[Department of Computer Science, Stony Brook University, NY, USA]{Huidong Liu}
\cortext[mycorrespondingauthor]{Corresponding author}
% \address{Radarweg 29, Amsterdam}
% \fntext[myfootnote]{Since 1880.}

% \author[mymainaddress,mysecondaryaddress]{Elsevier Inc}

% \author[mysecondaryaddress]{Global Customer Service\corref{mycorrespondingauthor}}
% \cortext[mycorrespondingauthor]{Corresponding author}

% \address[mymainaddress]{1600 John F Kennedy Boulevard, Philadelphia}
% \address[mysecondaryaddress]{360 Park Avenue South, New York}

\begin{abstract}
Well-labeled datasets of  chest radiographs (CXR) are often difficult to acquire due to the high cost of annotation. Thus, it is desirable to learn a robust and transferable representation in an unsupervised manner to benefit downstream tasks that lack annotated data. Unlike natural images, medical images often have their own domain prior;  e.g., we observe that many pulmonary diseases, such as the Coronavirus Disease 2019 (COVID-19), manifest as characteristic changes in the lung tissue texture rather than the anatomical structure. Therefore, we hypothesize that studying only the tissue texture without the influence of possible structure variations would be advantageous for downstream prognostic and predictive modeling tasks. In this paper, we propose a generative framework, the Lung Swapping Autoencoder (LSAE), that learns a factorized representation of a CXR to \textit{disentangle} the tissue texture representation from the anatomic structure representation. Specifically, by adversarial training, the LSAE is optimized to generate a hybrid image that preserves the lung shape in one image but inherits the lung texture of another image.
To demonstrate the effectiveness of the disentangled texture representation, we evaluate the texture encoder $Enc^t$ in LSAE on a large-scale thoracic disease dataset, ChestX-ray14 (N=112,120), and our own multi-institutional COVID-19 outcome prediction dataset, COVOC (N=340 (Subset-1) + 53 (Subset-2)).
On both datasets, we reach or surpass the state-of-the-art by finetuning the texture encoder in LSAE with a model that is 77$\%$ smaller than a baseline Inception v3 architecture. Additionally, in semi-supervised and linear evaluation settings with a similar model budget, the texture encoder in LSAE is also competitive with the state-of-the-art, MoCo. By ``re-mixing" the disentangled texture and shape factors, we generate meaningful hybrid images that can augment the training set. The proposed data augmentation method can further improve COVOC prediction performance. Moreover, the improvement is consistent even when we directly evaluate the Subset-1 trained model on Subset-2 without additional finetuning.
%On ChestX-ray14, $Enc^t$ achieves competitive results compared with the  state-of-the-art in both fully-supervised (79.0\%) and semi-supervised settings 
%(1\% labels: 61.3\%; 10\% labels: 73.2\%). On COVOC Subset-1, $Enc^t$ has 13\% reduced error with a model that is 77$\%$ smaller than a baseline Inception v3 architecture.
%In addition, using generative modeling, 
The code is available at:
\url{https://github.com/cvlab-stonybrook/LSAE}
\end{abstract}


\end{frontmatter}

% \linenumbers

\section{Introduction} \label{intro}
Chest radiographs (CXRs) are the primary imaging modality for front-line evaluation of suspected lung diseases, especially in the COVID-19 pandemic. They are widely available, portable, and have reduced risk of cross infection compared to Computed Tomography (CT).
Despite these advantages, CXRs are also less sensitive to subtle disease changes compared to CT scans [\cite{litmanovich_review_2020}]. For CXRs, it is often challenging to collect large-scale well-annotated datasets, especially for emerging diseases like COVID-19, due to the high cost of expert annotation. Therefore, it is important to develop algorithms that are able to learn a robust and transferable representation in an unsupervised/self-supervised way. Most self-supervised methods are designed for natural images and do not exploit any domain-specific priors that might exist for medical images.

COVID-19 manifests as infiltrates such as ground-glass opacities on CXR [\cite{litmanovich_review_2020}]. These findings are examples of changes in texture that appear as fuzzy-white areas within the lung fields.
% When observing recent COVID-19 CXRs, we found that lung tissue texture may change drastically during hospitalization due to varying infiltrate levels.
However, chest anatomy remains mostly unchanged throughout the disease course of an individual patient. Therefore, we hypothesize that \textit{disease information is more related to lung tissue texture rather than the anatomical structure of the lung}. 
To be concise, we use the terms \textit{texture} and \textit{structure} in the rest of the article.
% Our hypothesis is also supported by recent findings that COVID-19 on CXR are observed as opacities within lung regions, and their extent and location have been associated with severity of disease course \cite{toussie_clinical_2020,wong_frequency_2020}.

\begin{figure}
\centering
\includegraphics[width=\textwidth]{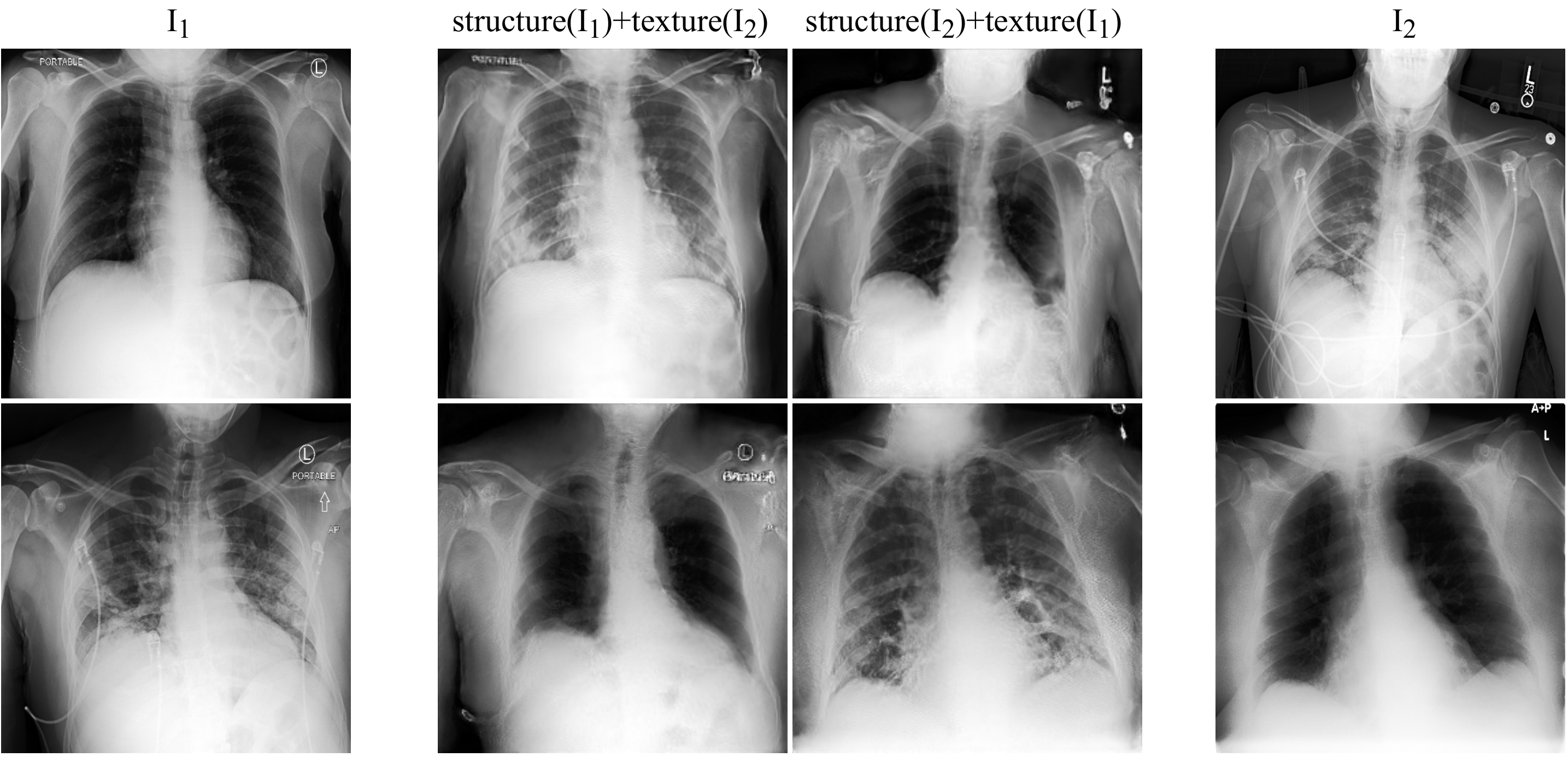}
\caption{\textbf{Lung Swapping Result.} Two examples of lung swapping between images in column $I_1$ and images in column $I_2$. The Lung Swapping Autoencoder (LSAE) is able to successfully transfer target lung textures  without affecting the lung shape. The swapping results are shown in the second and the third columns.}
\label{fig1}
\end{figure}

\begin{figure}[t]
\centering
\includegraphics[width=0.9\textwidth]{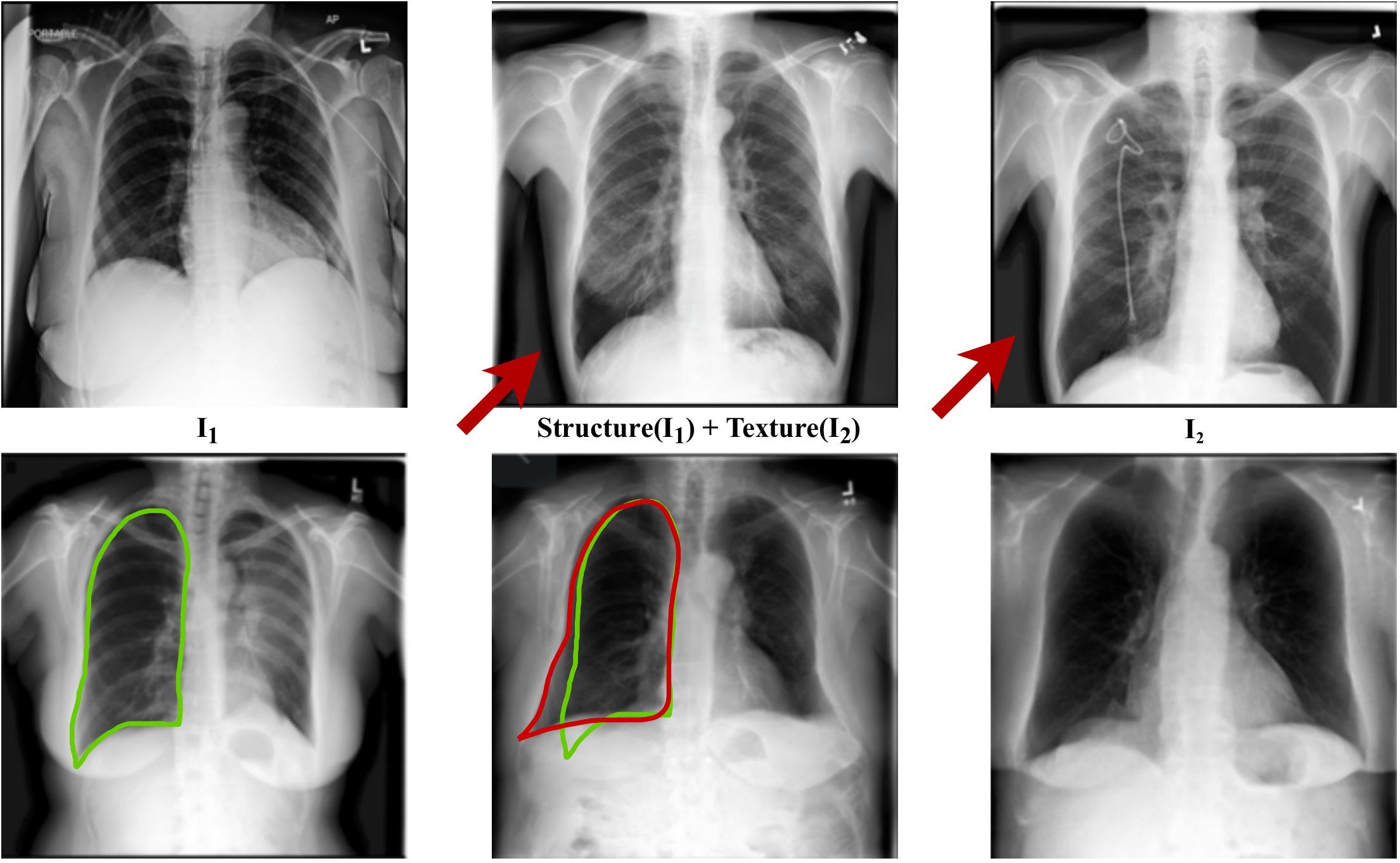}
\caption{\textbf{Two Problems in SAE.} The \textbf{first} problem is the irrelevant out-of-lung texture transfer as shown in the upper row. As indicated by the red arrows, these two out-of-lung regions show similar texture patterns which is undesired. The \textbf{second} problem is the lung shape distortion as shown in the lower row. The {\color{green} green} contour outlines the shape of lung in $I_1$ which should be preserved in the middle hybrid image. However, the lung shape in the middle hybrid image outlined in {\color{red} red} looks obviously distorted from the {\color{green} green} contour.}
\label{fig:problems_in_sae}
\end{figure}

Motivated by the above observation, we propose the Lung Swapping AutoEncoder (LSAE), that learns a factorized representation of a CXR to \textit{disentangle} the \textit{texture factor} from the \textit{structure factor}. LSAE shares the same core idea as the recently-proposed Swapping AutoEncoder (SAE) [\cite{park2020swapping}], namely that a successful disentanglement model should be able to generate a realistic hybrid image that merges the structure of one image with the texture of another. 
To achieve this goal, images are encoded as a combination of two latent codes representing structure and texture respectively. The SAE is trained to generate realistic images from the swapped codes of arbitrary image pairs in an adversarial learning manner. To supervise the synthesis of target texture, SAE is further forced to mimic the pattern of patches sampled from the target texture image.
However,  this vanilla SAE does not work well for CXR disentanglement (see Fig~\ref{fig:problems_in_sae}).
First, by sampling texture patches from the whole target image, irrelevant out-of-lung textures diminish the effect of the in-lung texture transfer (see the upper row in Fig~\ref{fig:problems_in_sae}).
Second, because texture supervision is derived from image patches, irrelevant structure clues may leak into the hybrid image, resulting in undesired lung shape distortion and interference with successful disentanglement (see the lower row in Fig~\ref{fig:problems_in_sae}).

% To address these two problems, 
We address the above two problems with two contributions. 
The first contribution is based on the observation that the infiltrates of interest are located within lung zones. Hence, sampling patches from the lung area instead of the whole image for texture supervision can alleviate the disturbance introduced by irrelevant texture patterns. The second contribution is applying a patchwise contrastive loss which explicitly forces out-of-lung local patches in the hybrid image to mimic the corresponding patch in the structure image. Thus, we can prevent structure information in texture patches from leaking into the hybrid image which can cause undesired lung shape distortion visually. 
We introduce this SAE equipped with the new designs as the Lung Swapping Autoencoder (LSAE). 
LSAE, trained on a large public CXR dataset, ChestX-ray14 [\cite{wang2017chestx}], can generate realistic, plausible hybrid CXRs with one patient's lung structure and another patient's disease texture (see Fig~\ref{fig1}. Please refer to Fig~\ref{more_results} for additional lung swapping examples).
% We evaluate such images quantitatively in Sec.~\ref{sec:results}.
We further demonstrate disentanglement performance quantitatively through disease texture distance, lung segmentation metrics, and radiologist feedback.

If textures do indeed represent disease infiltrates, we expect the texture encoder $Enc^t$ in LSAE to be discriminative in downstream CXR semantic tasks. We evaluate the learned representation on two datasets, the large-scale ChestX-ray14 database (N=112,120) and our multi-institutional COVID-19 outcome prediction dataset, COVOC (N=340(Subset-1)+53(Subset-2)). On ChestX-ray14, we compare $Enc^t$ in LSAE with state-of-the-art methods in linear evaluation, semi-supervised, and fully-supervised settings. Our method achieves competitive results with a model of almost half the size of ResNet18.
%$2 \times$ smaller model size (vs. ResNet18). 
Then we treat the $Enc^t$ as a pre-trained model on ChestX-ray14, and perform transfer learning to COVOC (Subset-1).
% When transferring to COVOC (Subset-1), 
$Enc^t$ in LSAE reduces the error rate by 13\% compared with a strong baseline Inception v3. Moreover, using LSAE, we generate hybrid images with textures and labels from COVOC training data and lung structures from  ChestX-ray14. Augmenting with these hybrid images further improve outcome (ventilation) prediction on COVOC. All experiments show that $Enc^t$ in LSAE learns effective and transferable representations of lung diseases.

We summarize our contributions as follows:
1) LSAE is the first approach to disentangle chest CXRs into structure and texture representations. LSAE achieves this complex task by leveraging explicit in-lung texture supervision and out-of-lung distortion suppression.
2) We achieve competitive performance compared with state-of-the-art approaches on the large-scale ChestX-ray14 dataset in both fully-supervised and semi-supervised settings for predicting pulmonary diseases.
3) We achieve superior performance for COVID-19 outcome prediction with an efficient model.
4) We propose a hybrid image augmentation technique to further improve clinical outcome prediction.

\section{Related Work}
\paragraph{Image-to-image Translation and Image Manipulation}
Image-to-image translation can be viewed as a conditional generation task. It takes a source image as input and transforms it into another target domain, such as a semantic image or manipulated image, as demonstrated in the following works [\cite{Isola_2017_CVPR,zhu2017unpaired,huang2018munit,choi2018stargan,liu2019few}]. These methods tackle the image-to-image translation task under different experimental settings, such as supervised, unsupervised, multi-modal, and few-shot conditions. We can categorize these approaches with respect to different architectures and loss functions. Most image-to-image models adopt U-Net or Autoencoder architectures. The key difference between the two genres lies in the skip connection. Equipped with skip connections, U-Net-based models can be optimized efficiently and accurately. However, Autoencoders have the advantage that a compact latent space is preserved. Specifically, an Autoencoder is usually composed of an encoder and a decoder. The encoder maps from the source image to a latent space, and the decoder maps from the latent space to the target image. 
As for loss functions, most methods adopt a reconstruction loss. To synthesize more realistic images, adversarial loss can sometimes be applied as well. Additionally, cycle-consistency loss [\cite{zhu2017unpaired}] is useful when no paired data is available.

To manipulate the image, the latent space of the Autoencoder needs to be disentangled and semantically meaningful. After that, by interpolating along a particular latent direction, the decoder can generate target/manipulated images. Our baseline model Swapping Autoencoder (SAE) [\cite{park2020swapping}] also follows the above routine for image manipulation. We will introduce SAE in the methodology section.

\paragraph{Self-supervised Representation Learning}
Current self-supervised representation learning methods primarily fall in two bins, contrastive learning and siamese non-contrastive learning. Generally, self-supervision originates from the invariance of representation with respect to augmentation transformations. For contrastive learning [\cite{chen2020simple}], a network is trained to attract features of views from one image but repulse views from different images. However, contrastive training is hungry for negative pairs which requires a very large batch to achieve a good result. To alleviate the heavy memory consumption burden, \cite{wu2018unsupervised} proposed a memory bank containing only feature vectors for a large batch of data. Alternatively, MoCo [\cite{he2020momentum,chen2020improved}] utilizes a queue of representation vectors produced by a momentum encoder. Recently, siamese networks have been proposed that can learn  effective representations without contrastive loss. Specifically, \cite{grill2020bootstrap} proposed a simple model called BYOL (Bootstrap Your Own Latent) which removes the negative pair repelling in training. Surprisingly, BYOL does not collapse to a trivial solution due to the absence of negative pairs. Instead, with a momentum encoder and the siamese architecture, BYOL achieves even better performance than contrastive counterparts. After BYOL, \cite{chen2021exploring} proposed a simpler version by removing momentum encoder with direct copying in BYOL, and proved empirically that a stop gradient with respect to the target branch made the siamese non-contrastive learning work particularly well. Moreover, \cite{tian2021understanding} analyzed these non-contrastive siamese frameworks from a theoretical view, justifying some key designs including the stop gradient and asymmetric predictor. Compared with contrastive learning, non-contrastive learning does not require a very large min-batch to work well. However, medical images have their own domain priors. For example, as mentioned above, CXRs for a patient have similar lung structures but different textures. All of the above methods are designed for natural images which do not consider the domain knowledge of medical images.

\paragraph{COVID-19 Imaging Studies}
Previous medical image analysis in COVID-19 has mostly focused on disease diagnosis [\cite{hu_role_2021,lopez-cabrera_current_2021}]. CT scan based models have done better in predicting COVID-19 disease outcomes compared to CXR-based models [\cite{chassagnon_ai-driven_2021,lopez-cabrera_current_2021,li_novel_2021,bae2020predicting}]. This is largely due to the lack of COVID-19 CXR datasets with relevant endpoints and the limited information that CXRs contain relative to CT scans. Several data augmentation techniques have been proposed for CT scans in the COVID-19 setting [\cite{li_novel_2021}], but CXR approaches have continued to rely on publicly sourced datasets [\cite{cohen2020covidProspective,lopez-cabrera_current_2021}]. These datasets tend to be homogeneous %have previously been criticized for homogeneity
within COVID-19 positive and negative classes, and often lack disease outcome labels (hospitalization, mechanical ventilation requirement, etc.) [\cite{lopez-cabrera_current_2021,cohen2020covidProspective}].
% \hll{Another version: In the aspect of methodology, Autoencoder and Generative Adversarial Networks \cite{goodfellow2014generative} are widely used for representation learning and data augmentation by encoding and image generation. However, previous work mainly focus on specific domains\cite{Isola_2017_CVPR,huang2018munit,choi2018stargan,zhu2017unpaired}, such as the class of cat, while the disease level in chest X-rays can vary in a continuous way.
%Known for generating high-quality images, 
% Generative Adversarial Networks (GANs) \cite{goodfellow2014generative} and (adversarially trained) Autoencoders~\cite{li2017generative} have been widely used for data augmentation, including   medical image applications \cite{hou2016patch,sandfort2019data,DBLP:journals/corr/abs-1807-10225}. However, standard GAN-based image generation methods~\cite{Isola_2017_CVPR,huang2018munit,choi2018stargan,zhu2017unpaired} are not suitable for CXR generation due to the lack of explicit structure supervision, which can lead to generating distorted shapes.

\section{Methodology}
First, we briefly review SAE in Section~\ref{sae}. Then, in Section~\ref{lsae}, we present our Lung Swapping Autoencoder (LSAE) model and emphasize our key designs. Lastly, in Section~\ref{aug}, we illustrate how to generate hybrid images for data augmentation.
\subsection{Swapping AutoEncoder (SAE)} \label{sae}
Recently, Swapping AutoEncoder (SAE) [\cite{park2020swapping}] is proposed for image manipulation by disentanglement. It consists of an encoder $Enc$ and a decoder $Dec$, where $Enc$ is composed of a structure branch $Enc^s$ and a texture branch $Enc^t$ to encode the input into structure and texture latent codes $z^s$ and  $z^t$ separately.

\subsubsection{Latent Code Swapping.}
To achieve the goal of disentanglement, SAE aims to generate realistic images from swapped latent codes of arbitrary image pairs.
Specifically, two sampled images $I_1$ and $I_2$ are first encoded as $(z_1^s, z_1^t)$ and $(z_2^s, z_2^t)$. Then, the latent codes are swapped to get a hybrid code $(z_1^s, z_2^t)$. Finally, the hybrid code is decoded to yield a hybrid image $I_{\texttt{hybrid}}$ which is expected to maintain the structure of $I_1$ but present the texture of $I_2$. We use $G$ to denote the composite generation process as
$$G(I_1, I_2) \myeq Dec(Enc^s(I_1), Enc^t(I_2))$$
To optimize the network, we try to minimize the reconstruction error and also take advantage of adversarial training by introducing an auxiliary discriminator. 
$$
\begin{aligned}
&\mathcal{L}_{\texttt{recon}} = \mathop{\mathbb{E}}_{I_1 \sim \mathbf{X}} \Vert G(I_1, I_1) - I_1\Vert_1 \\
&\mathcal{L}_{\texttt{G}} = \mathop{\mathbb{E}}_{{I_1, I_2} \sim \mathbf{X} }-\log\big(D\big(G(I_1, I_1)\big)\big) - \log\big(D\big(G(I_1, I_2)\big)\big)
\end{aligned}
$$
where $\mathbf{X}$ denotes the image training set, and $D$ is a discriminator [\cite{goodfellow2014generative}].
% to make the reconstructed image $G(I_1, I_1)$ and the hybrid image $G(I_1, I_2)$ realistic. 
Note that the complete GAN loss also includes a discriminator part. To be concise, we only show the generator part.

\subsubsection{Texture Supervision.}
To ensure $I_{\texttt{hybrid}}$ texture matches that of $I_2$, SAE samples patches from $I_2$ to supervise the $I_{\texttt{hybrid}}$ texture. Another auxiliary network, patch discriminator $D_{\texttt{patch}}$, is trained to distinguish patches in $I_{\texttt{hybrid}}$ from patches in $I_2$. Adversarially, SAE is trained to generate a $I_{\texttt{hybrid}}$ whose patches can confuse $D_{\texttt{patch}}$ by mimicking the texture of $I_2$. The texture loss is formulated as $$\mathcal{L}_\texttt{tex} = \mathbb{E}_{\substack{\tau_1, \tau_2 \sim \mathcal{T} \\ I_1, I_2 \sim \mathbf{X}}} \big[-\log\big(D_{\texttt{patch}}\big(\tau_1\big(I_2\big), \tau_2\big(G(I_1, I_2)\big)\big)\big)\big]$$
where $\mathcal{T}$ is the distribution of multi-scale random cropping, and $\tau_1$ and $\tau_2$ are two random operations sampled from $\mathcal{T}$.

\subsection{Lung Swapping AutoEncoder (LSAE)} \label{lsae}
\begin{figure}[t]
\centering
\includegraphics[width=0.87\textwidth]{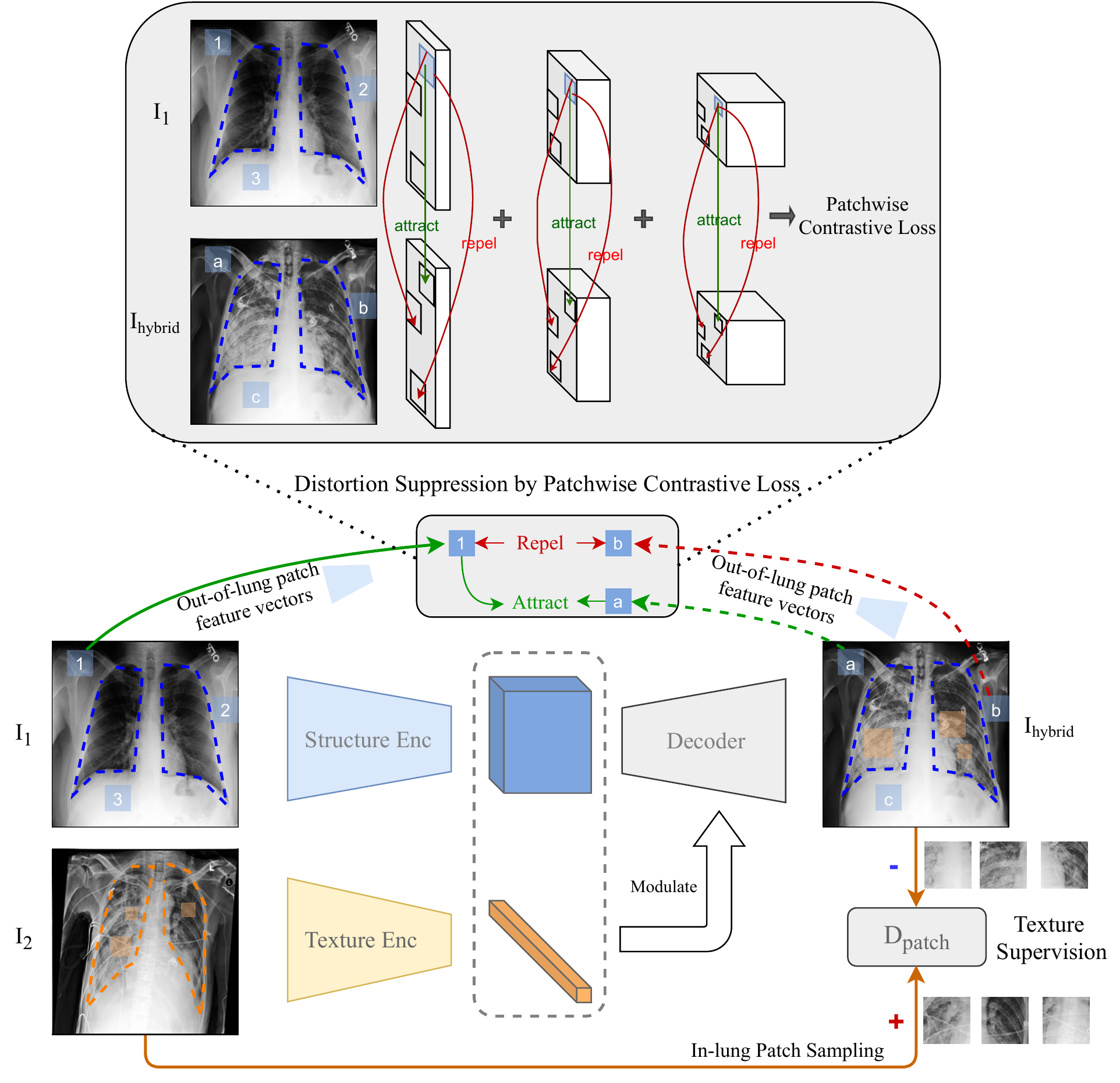} 
\caption{\textbf{Lung Swapping AutoEncoder} (LSAE) consists of a structure encoder, a texture encoder, and a decoder. After inputting  images $(I_1, I_2)$, the LSAE generates a hybrid image with factorized representations from $I_1$ and $I_2$ respectively. To ensure the texture in $I_{\texttt{hybrid}}$ matches $I_2$, we adversarially train a patch discriminator $D_{\texttt{patch}}$ to supervise texture synthesis within lungs. To ensure $I_{\texttt{hybrid}}$ maintains the structure of $I_1$, we apply a patchwise contrastive loss outside the lungs to minimize structural distortion.} \label{fig2}
\end{figure}

Using SAE directly to generate a hybrid CXR from an image pair does not work as desired. First, the disease level in $I_{\texttt{hybrid}}$ is usually diminished when compared with $I_2$. We hypothesize that the out-of-lung irrelevant texture patterns may hamper target texture synthesis in $I_{\texttt{hybrid}}$. %Later quantitative 
Results in Table~\ref{tab:swap} support our analysis. Please refer to the experimental part for detailed discussion on the results.
Second, since there is no structure supervision in SAE, $I_{\texttt{hybrid}}$ can show undesired lung shape distortion towards $I_2$ which indicates a suboptimal disentanglement.
These problems lead us to design two new strategies, in-lung texture supervision and out-of-lung structural distortion suppression. We name the SAE model equipped with the new designed strategies the \textit{Lung Swapping AutoEncoder} (LSAE).
\subsubsection{In-lung Texture Supervision.}
As infiltrates on CXRs are observed inside lung regions, we should cast the focus of texture synthesis in the in-lung region. Hence, we can adapt SAE by sampling texture patches from within the lung zone instead of the whole image. Guided by a lung segmentation mask which can be computed by a pre-trained model, we rewrite the texture supervision loss as,
\begin{equation}
\mathcal{L}_\texttt{inTex} = \mathop{\mathbb{E}}_{\substack{\tau_1, \tau_2 \sim \mathcal{T}_{\texttt{LungMask}}^{\texttt{in}} \\ I_1, I_2 \sim \mathbf{X}}} \bigg[-\log\bigg(D_{\texttt{patch}}\Big(\tau_1\big(I_2\big), \tau_2\big(G(I_1, I_2)\big)\Big)\bigg)\bigg]
\end{equation}
where most items are the same as $\mathcal{L}_\texttt{tex}$ except $\mathcal{T}$ in $\mathcal{L}_\texttt{tex}$, which is now replaced by $\mathcal{T}_{\texttt{LungMask}}^{\texttt{in}}$,  indicating that we only sample patches from in-lung regions to supervise the synthesis of target texture in $I_2$.
\subsubsection{Out-of-lung Structural Distortion Suppression.}
To suppress the undesired structural distortion, we further introduce a patchwise contrastive loss [\cite{park2020cut}] in LSAE. The patchwise contrastive loss, based on Noise Contrastive Estimation [\cite{gutmann2010noise}], is designed to preserve image content in image-to-image translation tasks. Hence, we adapt the patchwise contrastive loss to help our task by applying it in the following manner: We first encode $I_{\texttt{hybrid}}$ with structure encoder $Enc^s$. Then, we randomly sample a feature vector $q_i$ from the $h^{th}$ layer's output $Enc^s_h(I_{\texttt{hybrid}})$, and a bag of feature vectors $\{p^+, p^-_1, p^-_2, \cdots, p^-_{N-1} \}$ from $Enc^s_h(I_1)$, where $p^+$ is from the corresponding position with $q_i$, and $\{p^-\}$  from other positions. The loss objective is to force $q_i$ from $I_{\texttt{hybrid}}$ to attract to the corresponding feature vector $p^+$ but repel  the others $\{p^-\}_{N-1}$ from $I_1$. Therefore, we write it as a cross-entropy loss by regarding the process as an $N$-way classification problem where $q_i$ is the query and $p^+$ is the target:
% As the loss inherits the key idea of Noise Contrastive Estimation \cite{gutmann2010noise}, we name it as $l_{\texttt{NCE}}$. Here is the formula,
\begin{equation} \label{eq:nce}
l_{\texttt{NCE}}(q_i, \{p\}_N) = -\log \left(\frac{\exp(q_i \cdot p^+ / \alpha)}{\exp(q_i \cdot p^+ / \alpha) + \sum_{j=1}^{N-1}\exp(q_i \cdot p^-_j / \alpha)}\right)
\end{equation}
where $\alpha$ is the temperature, and $\{p\}_N = \{p^+\}\cup \{p^-\}_{N-1}$. Since we found experimentally that applying $l_{\texttt{NCE}}$ to the whole spatial region can disturb the texture synthesis within lungs, we only apply it outside the lung region. Thus, the final structural distortion suppression loss can be expressed as
\begin{equation}
\mathcal{L}_{\texttt{sup}} = \mathop{\mathbb{E}}_{\substack{\{\tau\}_N \sim \mathcal{T}^{\texttt{out}}_{\texttt{LungMask}} \\ I_1, I_2 \in \mathbf{X}}} \sum_{h}^{H} \sum_{i}^{N} l_{\texttt{NCE}}\Big(\tau_i\big(Enc^s_h(I_{\texttt{hybrid}})), \{\tau\big(Enc^s_h(I_1)\big)\}_N\Big)
\end{equation}
where $\mathcal{T}^\texttt{out}_\texttt{LungMask}$ is the distribution of position sampling outside the lung region guided by the mask, and $H$ is the number of layers we apply the loss to.

\subsubsection{The Overall Objective.}
% After introducing all the core features in LSAE, we assemble them together to formulate our final optimization objective as follows,
Combining the core designs in LSAE, we term the final loss as
\begin{equation} \label{eq:hyper}
\mathcal{L} = \mathcal{L}_{\texttt{recon}} + \lambda_1 \mathcal{L}_{\texttt{G}} + \lambda_2 \mathcal{L}_{\texttt{inTex}} + \lambda_3 \mathcal{L}_{\texttt{sup}}
\end{equation}
where $\lambda_1$, $\lambda_2$, and $\lambda_3$ are tunable hyper-parameters. The selected hyper-parameters in experiments are provided further in Sec.~\ref{sec:imple}.
\subsection{Hybrid Image Augmentation} \label{aug}
If the texture in $I_2$ can be transferred to $I_{\texttt{hybrid}}$, we assume the label (e.g., ventilation) of $I_2$ is also attached to $I_{\texttt{hybrid}}$, even if the lung structure is different. Based on this hypothesis, we design a new method for data augmentation: Given an image $I^{\texttt{dst}}$ in the target domain, e.g., COVID-19 Outcome prediction dataset, COVOC, in our context, we sample $K$ images $\{I^\texttt{{src}}\}$ from a source domain, e.g., ChestX-ray14. Then, we use LSAE to take $I^{\texttt{dst}}$ as the texture template and images from $\{I^\texttt{{src}}\}$ as structure templates to generate $K$ hybrid images $\{I_{\texttt{hybrid}}\}$. We label $\{I_{\texttt{hybrid}}\}$ with the label of $I^{\texttt{dst}}$. Following this protocol, we can enlarge the training set in the target domain to $(K+1)$ times of origial size. Experimental results show that training with the additional generated hybrid images can improve performance further on the COVID-19 ventilation prediction task.

\section{Experimental Design and Results}
\subsection{Dataset Description} \label{data}
\noindent \textbf{ChestX-ray14} [\cite{wang2017chestx}] is a large-scale CXR database consisting of 112,120 frontal-view CXRs from 32,717 patients. We report all the results based on the official split which consists of training ($\sim70\% $), validation ($\sim10\%$), and testing ($\sim20\%$) sets. Images from the same patient will only appear in one of the sets.

\noindent \textbf{COVID-19 Outcome (COVOC)} is a COVID-19 CXR dataset curated from three institutes. Each CXR in COVOC is labeled based upon whether the patient required mechanical ventilation (henceforth ventilation) or not. Specifically, we split COVOC into two subsets. Subset-1 constitutes of 340 CXRs from 327 COVID-19 patients acquired in the first two institutes upon disease presentation. We further separate this subset randomly into 3 splits. Each split has 250 samples for training, 30 for validation, and 60 for testing; Subset-2 are composed of 53 images from the third institute. We only use this subset for testing in order to demonstrate the effectiveness of our method when generalizing to a new domain without adapting. As for positive/negative cases distribution, we list the statistics in Table~\ref{tab:covoc_dist}.

\begin{table}[t]
\centering
\begin{tabular}{c|ccc}
\hline
 & \makecell{Subset-1 \\ Train/Val/Test/Subtotal} & \makecell{Subset-2 \\ Subtotal(Test)} & Total \\
\toprule
positive & 81/10/20/111 & 27 & 138 \\
\hline
negative & 169/20/40/229 & 26 & 255 \\
\hline
\end{tabular}
\caption{\textbf{Class-wise data distribution on COVOC.} for prediction of ventilation requirement. There are two subsets in COVOC. Subset-1 is used to train, validate, and test models. Subset-2 is reserved only for evaluating the generalizing ability of models (usually trained on Subset-1 training set) to a new domain without any adaptation.}
\label{tab:covoc_dist}
\end{table}

\subsection{Implementation} \label{sec:imple}
\noindent \textbf{Architecture}. In LSAE, both decoder and discriminator architectures follow StyleGAN2. Encoders are built with residual blocks [\cite{he2016deep}]. Among them, the texture encoder outputs a flattened vector while structure encoder's output preserves spatial dimension. \\
\noindent \textbf{Preprocessing}. The input image size of both encoder and discriminator is $256 \times 256$. Patch sizes sampled for texture supervision vary from $16 \times 16$ to $64 \times 64$. We also randomly rotate the cropped patches up to 60 degrees. We preprocess all CXRs with histogram equalization. The lung masks are pre-computed by a pre-trained segmentation model from [\cite{konwer2021predicting}].
\\
\noindent \textbf{Hyper-parameters}. 
The temperature $\alpha$ in Eq.~\ref{eq:nce} is set to 0.07. We set $\lambda_1 = 0.5$, $\lambda_2 = 1$, and $\lambda_3 = 1$ in Eq.~\ref{eq:hyper}.

\begin{table}[h!]
\centering
\begin{tabular}{cc}
\hline
Questions & Images \\
\toprule
\makecell{Q1: Is this a real \\ or a fake image?} & \includegraphics[scale=0.2]{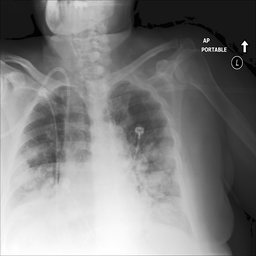} \\
\hline
\makecell{Q2: Is this patch cropped \\ from a real image?}
& \includegraphics[scale=0.32]{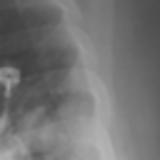} \\
\hline
\makecell{Q3: Which CXR (A or B) \\ has a more similar disease \\ level to the reference one?} & \includegraphics[scale=0.25]{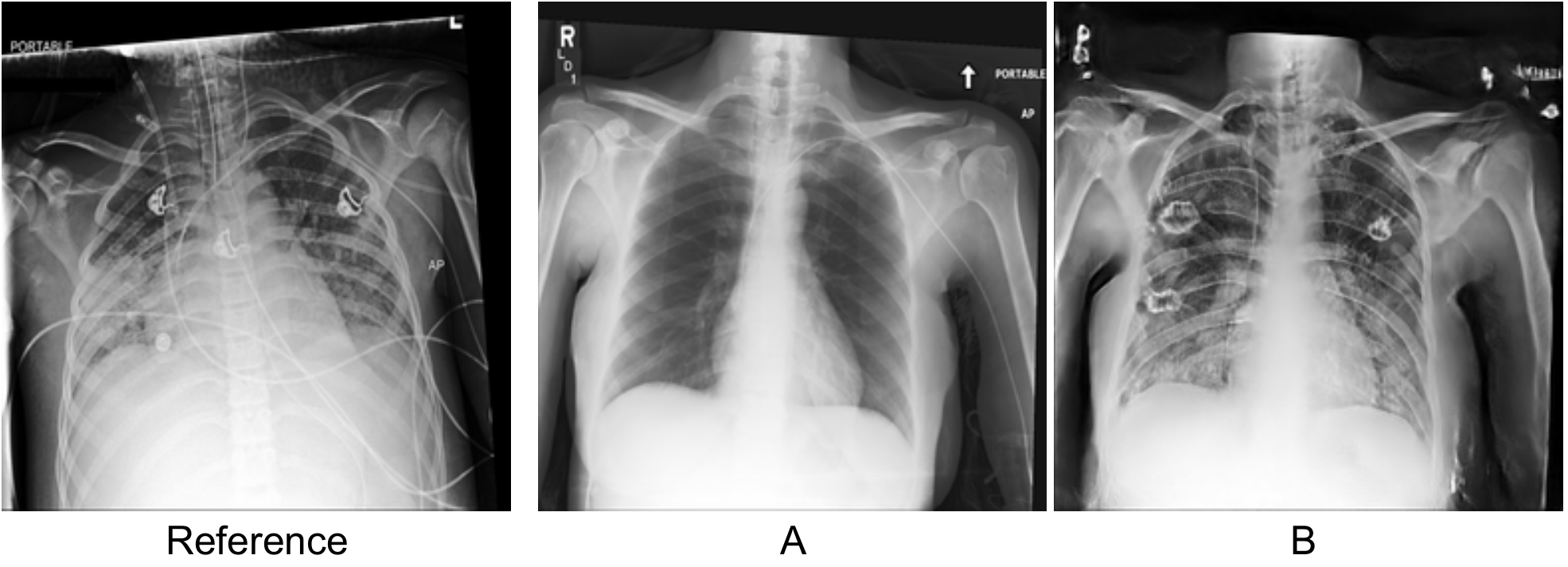} \\
\hline
\makecell{Q4: Which CXRs \\
do you think have \\
fibrosis/pneumonia/infiltration?} & \includegraphics[scale=0.3]{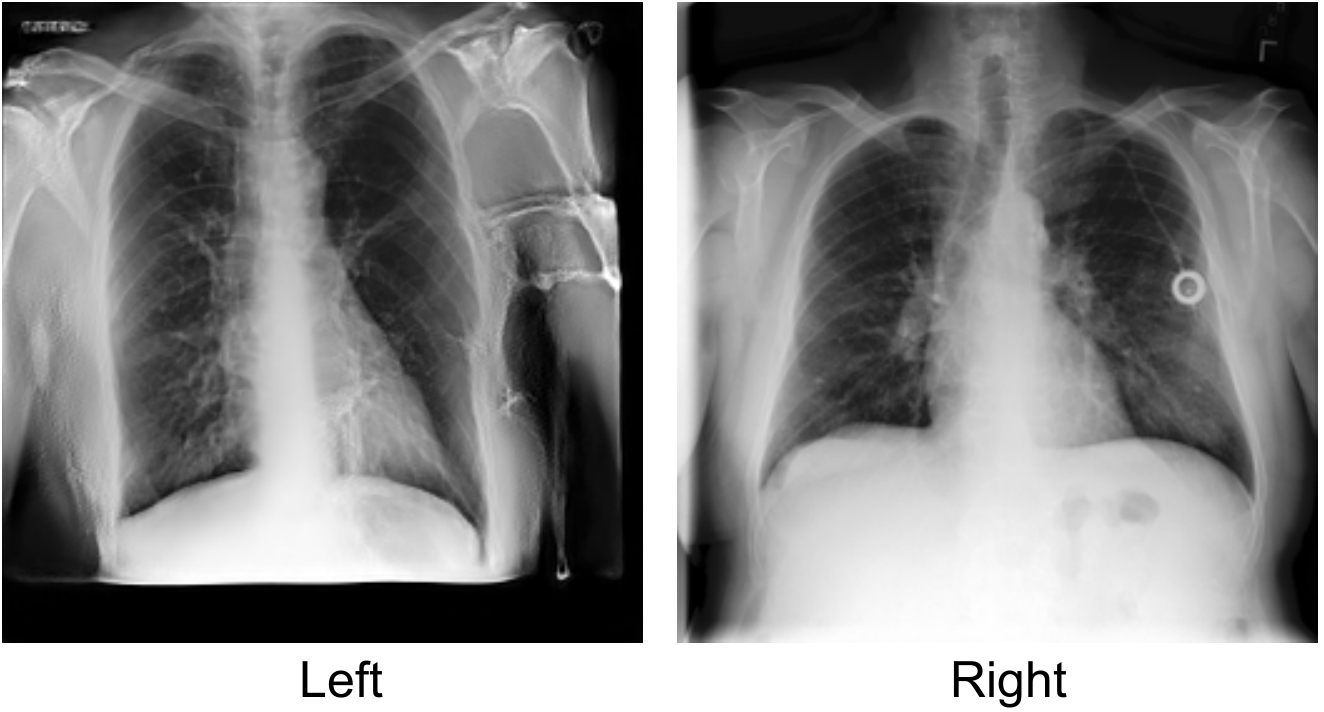} \\
\hline
\end{tabular}
\caption{\textbf{Survey Questions}. The first two questions seek to assess the image synthesis quality in both the image and patch level. The third question is designed to verify the correctness of the proposed Masked SIFID by measuring the matching rate between radiologists' judgements and Masked SIFID's measurements. The last question aims to prove whether LSAE can transfer the target disease texture correctly.
%In the first two questions regarding  image quality, radiologists misconstrued 56$\%$ of the generated images as real, and 74$\%$ of the hybrid image patches as real. The third question is to verify the correlation between Masked SIFID and disease level distance. When picking which of two query images is closer  to the reference, 78.67$\%$ of the radiologist answers match with the Masked SIFID. The fourth question is to ascertain whether our method can transfer disease correctly. When the radiologists were showm the original and the hybrid image with the same texture, 60$\%$ of the images passed the test.
}
\label{tab:survey_questions}
\end{table}

\begin{table}[t]
    \centering
    \begin{tabular}{cc}
        \includegraphics[trim=27pt 0 14pt 0, width=0.46\linewidth]{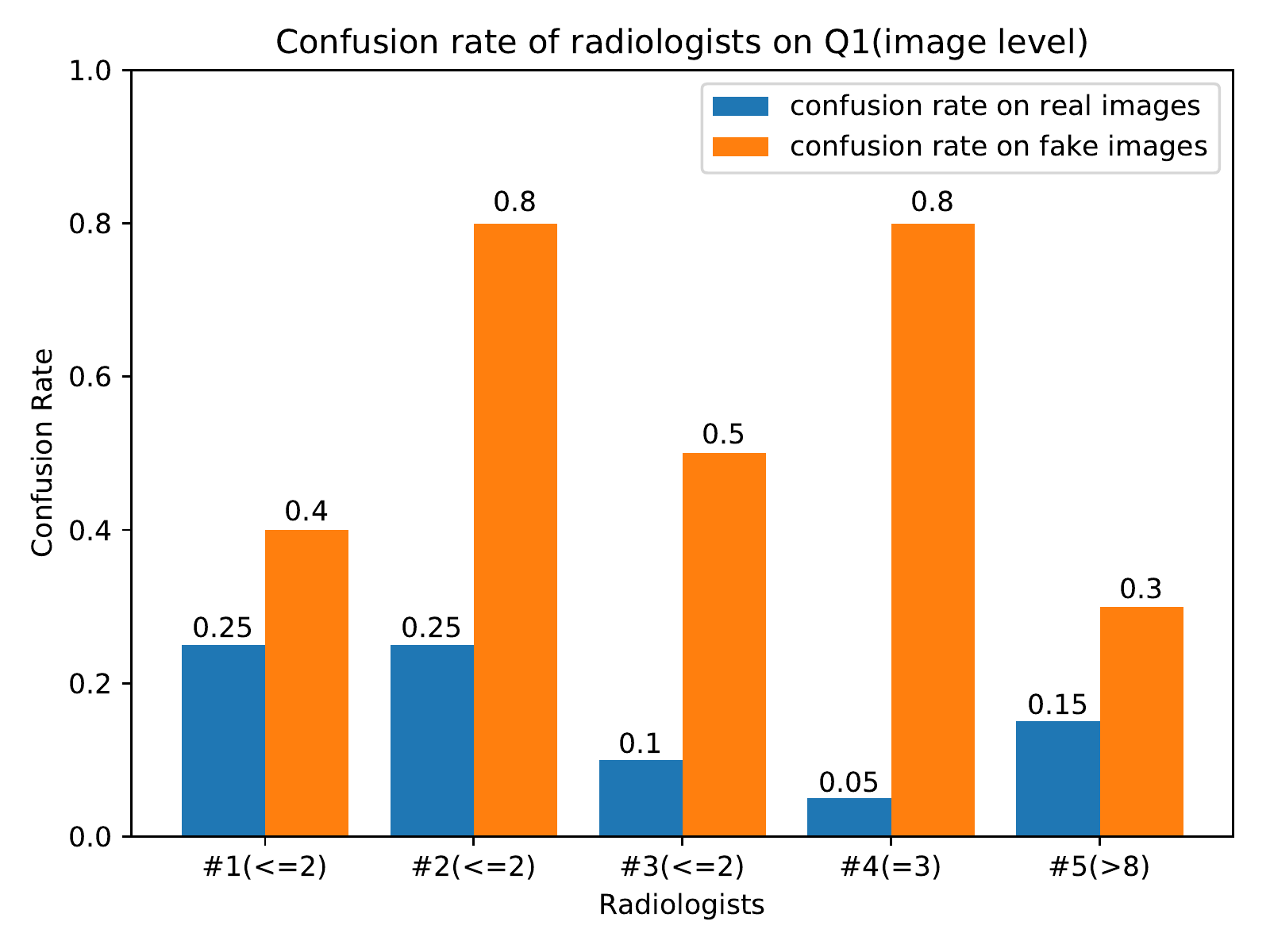} & \includegraphics[trim=27pt 0 14pt 0, width=0.46\linewidth]{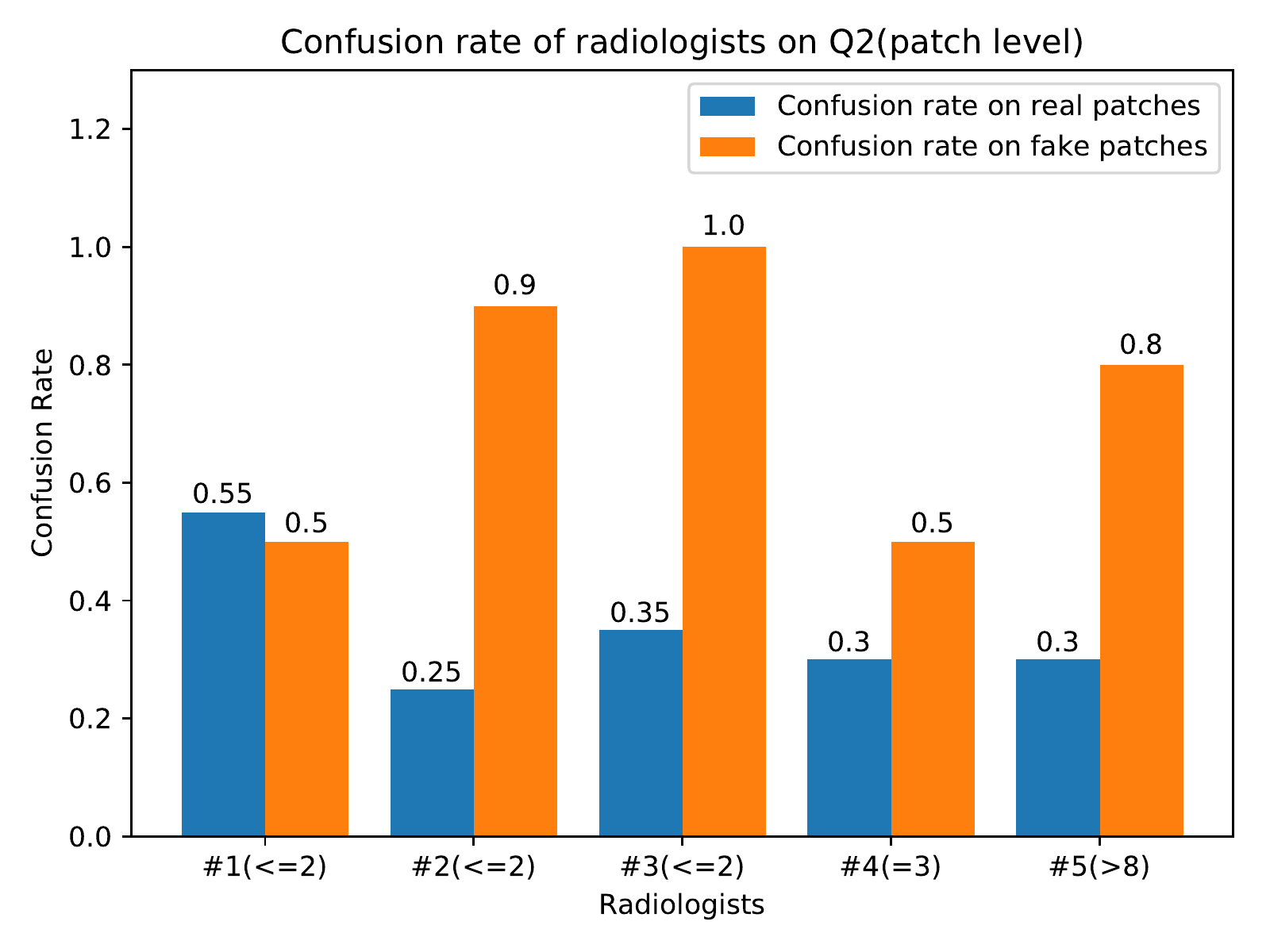} \\
         (a) & (b) \\
        % Q1 results & Q2 results \\
        \includegraphics[trim=27pt 0 14pt 0, width=0.46\linewidth]{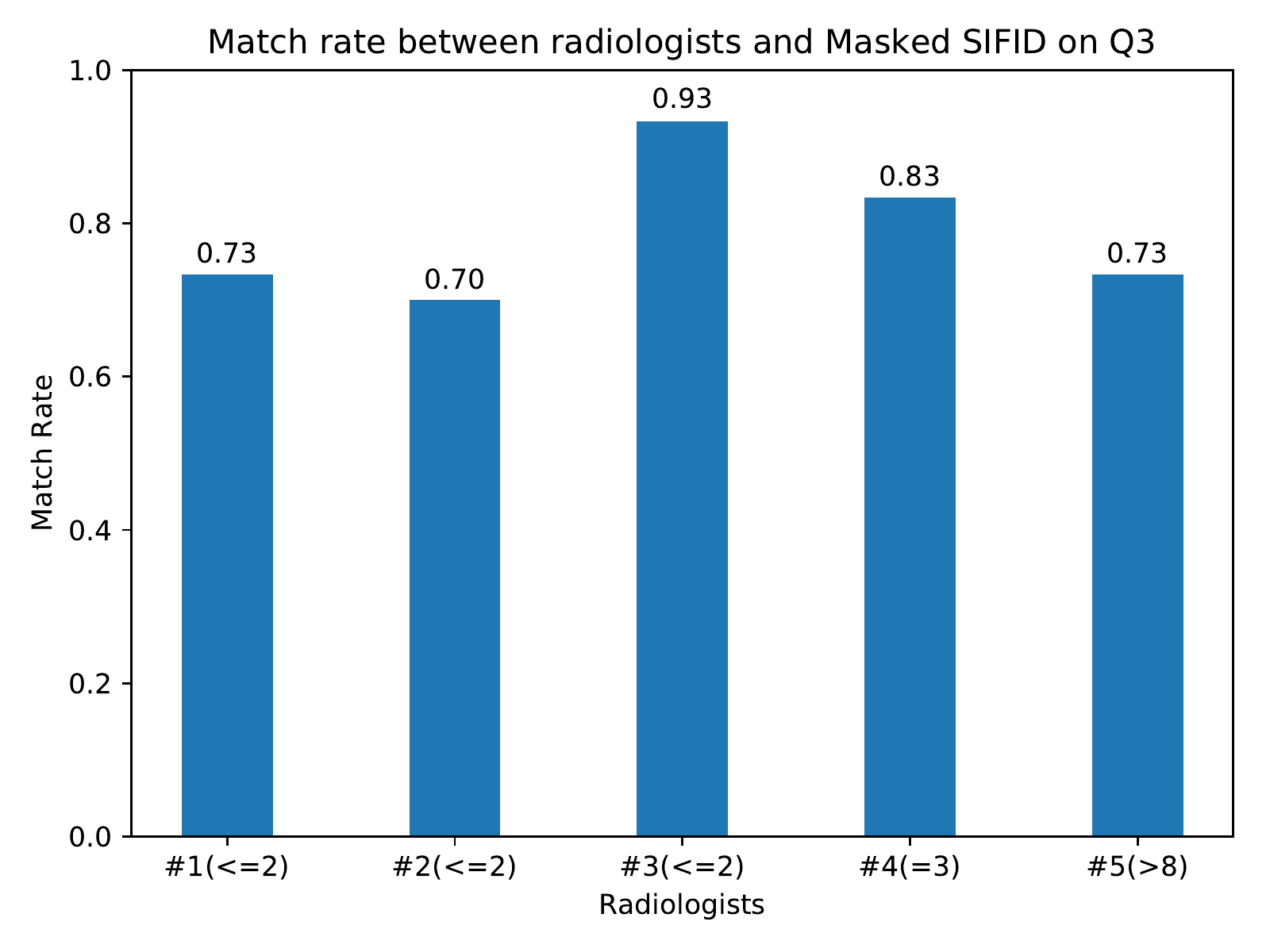} & \includegraphics[trim=27pt 0 14pt 0, width=0.46\linewidth]{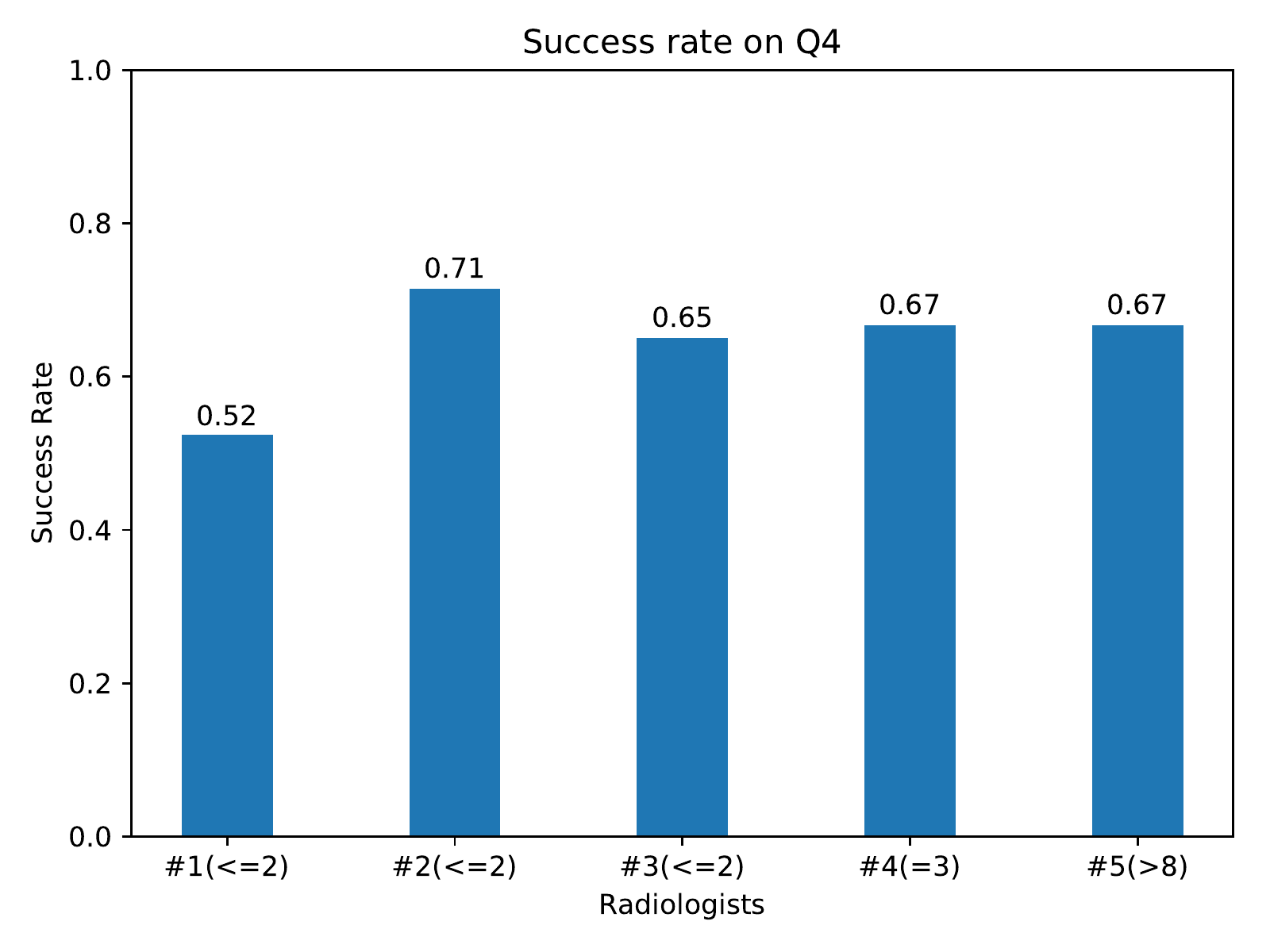} \\
        (c) & (d)
        % \\
        % Q3 results & Q4 results
    \end{tabular}
    \caption{\textbf{Radiologists' Survey Results}. The x-axes of all the plots denote indices of radiologists with the corresponding years of radiology training. (a), (b), (c), and (d) show the results for Q1, Q2, Q3, and Q4, respectively. In the upper row, we also list the confusion rates on real images for reference. The match rate in lower-left plot is computed by $\frac{\#QsHaveSameAnswers}{\#Qs}$. The success rate in lower-right plot is computed by $\frac{\#QsBothHaveDisease}{\#QsBothHaveDisease+\#QsOnlyRefHasDisease}$. The final averaged results are reported
    in Sec.~\ref{sec:results}.}
    \label{fig:survey}
\end{table}

\noindent \textbf{Training}. LSAE is optimized by Adam with a learning rate of 1e-3. Each training batch contains 16 images that are distributed in 2 NVIDIA GeForce RTX 2080 Ti GPUs. To regularize both discriminators, we apply R1 regularization every 16 iterations. When finetuning the texture encoder on COVOC, we add a randomly-initialized linear layer for prediction, and set the batch size to 56. The texture encoder is still optimized by Adam. We adopt an exponential decay schedule for learning rate which starts from 0.004 and ends at 0.001. Our code is developed based on [\cite{sae-pytorch}] and runs on PyTorch 1.7. We use Wandb [\cite{wandb}] to monitor the training process.

\subsection{Hybrid CXR Generation}
Our first experiments utilize the large ChestX-ray14 dataset to pre-train LSAE by learning to generate hybrid CXRs from swapped latent code in an unsupervised way.

\subsubsection{Experimental Settings.} We train LSAE on the training set of ChestX-ray14 with a batch of 16 images for 150K iterations. To evaluate the performance, we define a lung swapping task as follows, First, we create two sets of 9,000 images, $\{I^+\}$ and $\{I^-\}$, from the test set of ChestX-ray14. $\{I^+\}$ are sampled from the images diagnosed with at least one of the diseases: infiltration, pneumonia, and fibrosis, have been described to share similar imaging findings with COVID-19 [\cite{salehi_coronavirus_2020}]. In contrast, $\{I^-\}$ are sampled from healthy lungs. We pair each image in $\{I^+\}$ with an image in $\{I^-\}$ in a random order. Thus, we can now generate two hybrid images by mixing texture and structure for each image pair in both directions, i.e., $G(I^+, I^-)$ and $G(I^-, I^+)$. We will measure the distance of the set of hybrid images between both the target texture image set and the structure image set.

\begin{figure}[t]
\centering
\includegraphics[width=1\textwidth]{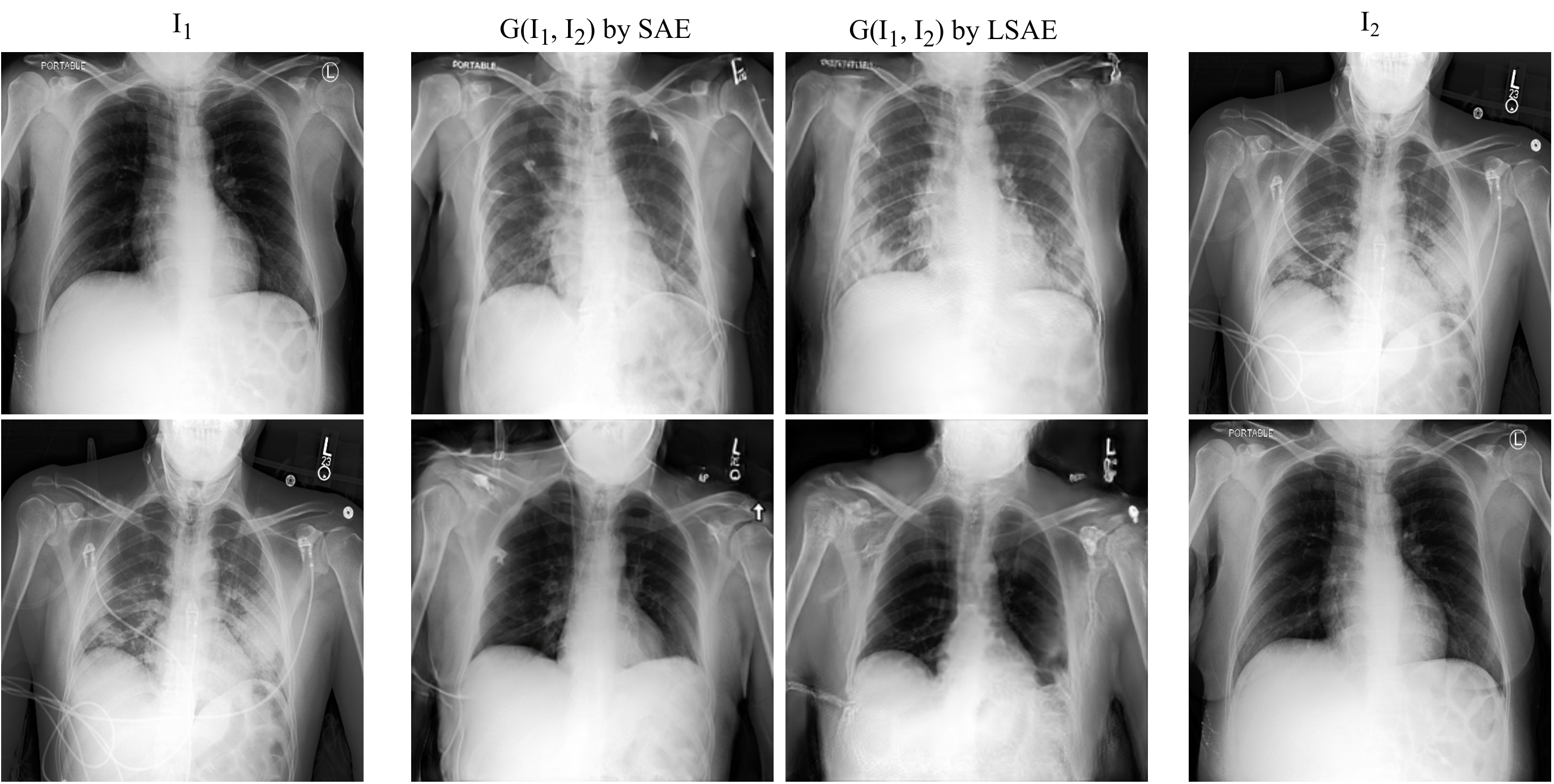} 
\caption{\textbf{Comparison with SAE.} As indicated in Table~\ref{tab:swap}, LSAE can suppress lung shape distortion while synthesizing target texture. Here we compare LSAE with SAE in a qualitative manner. Similar to Fig.~\ref{fig1}, the first and last columns are real images for mixing. The second column is the hybrid image generated by the baseline SAE while the third column is generated by our LSAE. It is obvious that SAE suffers from shape distortion while LSAE can suppress the undesired distortion effectively.} \label{fig:sae}
\end{figure}

\subsubsection{Evaluation Protocol}
% We evaluate the above swapping task in three ways. 
\noindent \textbf{Quantitative evaluation.} First, to measure the disease level distance between $I_{\texttt{hybrid}}$ and $I_2$, we propose a new metric, Masked SIFID. Masked SIFID is designed based on SIFID [\cite{rottshaham2019singan}] which calculates the FID distance between two images in the Inception v3 feature space. To customize it to disease level distance, we only consider features within the lung region. Additionally, we use the ChestX-ray14 pre-trained Inception v3 (mAUC: 79.56$\%$) to infer features.
Second, to quantify structural distortion, we use lung segmentation metrics as surrogates. Given a lung segmentation model $Seg$ [\cite{konwer2021predicting}], we can compute segmentation metrics by treating $Seg(G(I_1, I_2))$ as the query and $Seg(I_1)$ as the ground truth.

\noindent \textbf{Qualitative evaluation.} To evaluate the results qualitatively, we solicited feedback from 5 radiologists through a 4-question survey. The questions are shown in Table.~\ref{tab:survey_questions}. In particular, the first two questions seek to assess the image synthesis quality in both the image and patch level. The third question is designed to verify the correctness of the proposed Masked SIFID. The last question aims to prove whether LSAE can transfer the target disease texture correctly.

\begin{table}[t]
\centering
\begin{tabular}{c!{\vrule width 2pt}c|ccc}
\hline
Method & Masked SIFID $\downarrow$ & mIoU $\uparrow$ & Pixel Acc $\uparrow$ & Dice $\uparrow$ \\
\toprule
Init & 0.0335 & 0.60 & 0.82 & 0.72 \\
SAE & 0.0257 & 0.76 & 0.91 & 0.85 \\
LSAE & \textbf{0.0245} & \textbf{0.91} & \textbf{0.97} & \textbf{0.95} \\
\hline
\end{tabular}
\caption{\textbf{Evaluation of Hybrid Image Generation}. We report Masked SIFID for evaluation of disease transgfer. To evaluate distortion suppression, we report mIoU, pixel accuracy, and Dice. LSAE surpasses SAE in both texture synthesis and structure maintenance. We report the average results of texture transfer over two directions, i.e., texture transfer from $\{I^+\}$ to $\{I^-\}$, and from $\{I^-\}$ to $\{I^+\}$.}  \label{tab:swap}
\end{table}

\begin{table}[t]
\centering
\begin{tabular}{c|ccc}
\hline
\makecell{Pre-train} & Method & Params & mAUC $\uparrow$ \\
\toprule
\multirow{6}{*}{Supervised} & CXR14-R50 [\cite{wang2017chestx}] & 23M & 0.745 \\
&ChestNet [\cite{wang2018chestnet}] & 60M & 0.781 \\
&CheXNet$^\ast$ [\cite{rajpurkar2017chexnet}] & 7M & 0.789 \\
&Inception v3 & 22M & \textbf{0.796} \\
\hline
\multirow{2}{*}{Unsupervised} & MoCo-R18 [\cite{he2020momentum}] & 11M & 0.786 \\
& $Enc^t$ in LSAE & \textbf{5M} & \textbf{0.790} \\
\hline
\end{tabular}
\caption{\textbf{Performance comparison on ChestX-ray14}. The texture encoder in LSAE achieves competitive results with a smaller model size. $^\ast$Note that the data split in [\cite{rajpurkar2017chexnet}] is not released. We reimplemented CheXNet on the official split. Note that CheXNet follows the structure of DenseNet-121. The inception v3 model is also self-implemented.}  \label{tab:cxr14}
\end{table}

% Please see supplementary material for survey details.

\begin{table}[t]
\centering
\begin{tabular}{ccc}
\hline
Lung Diseases & ChestNet [\cite{wang2018chestnet}] & $Enc^t$ in LSAE \\
\midrule
Atelectasis	& 0.7433 & \textbf{0.7545} \\
Cardiomegaly & \textbf{0.8748} & 0.8668 \\
Effusion & 0.8114 & \textbf{0.8214} \\
Infiltration & 0.6772 & \textbf{0.6927} \\
Mass & 0.7833 & \textbf{0.7948} \\
Nodule & 0.6975 & \textbf{0.7380} \\
Pneumonia & 0.6959 & \textbf{0.7088} \\
Pneumothorax & 0.8098 & \textbf{0.8442} \\
Consolidation & 0.7256 & \textbf{0.7297} \\
Edema & 0.8327 & \textbf{0.8401} \\
Emphysema & 0.8222 & \textbf{0.8547} \\
Fibrosis & 0.8041 & \textbf{0.8078} \\
Pleural Thickening & \textbf{0.7513} & 0.7489 \\
Hernia & \textbf{0.8996} & 0.8667 \\
\hline
mAUC & 0.781 & \textbf{0.7906} \\
\hline
\end{tabular}
\caption{\textbf{Class-wise classification performance on  ChestX-ray14}. We report class-wise AUC metric of our model, i.e., the texture encoder in a pre-trained LSAE. The mean AUC corresponds to the results reported in Table.~\ref{tab:cxr14}. We also list class-wise AUC of ChestNet for a reference.} \label{tab:cxr14-c}
\end{table}

\subsubsection{Results} \label{sec:results}
\noindent \textbf{Disease Transfer}. In Table~\ref{tab:swap}, to get reference values, we first report initial results by directly comparing $\{I^-\}$ with $\{I^+\}$. LSAE achieves Masked SIFID of 0.0245 when compared with 0.0257 using SAE and the initial distance 0.0335, demonstrating that our design of in-lung texture supervision works as expected. 

\noindent \textbf{Distortion Suppression}. LSAE also outperforms SAE by an average of 12.7\% in all segmentation metrics, achieving over 90$\%$ in all segmentation metrics, which proves out-of-lung patchwise contrastive loss can effectively suppress structural distortion. In Fig.~\ref{fig:sae}, we visualize the image swapping results of both the baseline SAE and our LSAE. It can be observed qualitatively that LSAE does perform better with respect to the suppression of lung shape distortion.

\noindent \textbf{Radiologists' Survey}. The results of radiologists' survey are shown in Fig.~\ref{fig:survey}. The x-axes of all the plots denote indices of radiologists with the corresponding years of residency. Respectively, (a), (b), (c), and (d) sub-figures shows the results for Q1, Q2, Q3, and Q4. In the first two questions regarding image quality, radiologists misconstrued 56$\%$ of the generated images as real, and 74$\%$ of the hybrid image patches as real on average. To get a reference, we also list the confusion rates of real images in the upper row of Fig.~\ref{fig:survey} as blue bars. The third question attempted to verify a correlation between Masked SIFID and disease level distance. After the radiologists selected the query image closer to the reference, we compute the proportion of radiologists' answers matching with Masked SIFID. Specifically, match rate is computed by  $\frac{\#QsHaveSameAnswers}{\#Qs}$. On average, the match rate in Q3 is 78.67$\%$. This justifies the proposed Masked SIFID for evaluating disease transfer. The fourth question was to ascertain whether our method can transfer disease correctly. When the radiologists were shown the original image and the hybrid image generated with the same texture, we compute the success rate by $\frac{\#QsBothHaveDisease}{\#QsBothHaveDisease+\#QsOnlyRefHasDisease}$. Specifically, the success rate of 60$\%$ in Q4 means 60$\%$ of the hybrid images are identified to have the same lung disease as the original texture image on average.

\subsection{Semantic Prediction in CXRs by Texture Encoder} \label{exp:pred}
Based on our hypothesis that pulmonary diseases are tightly related to CXR texture, the texture encoder in a well-trained LSAE should be discriminative on CXR semantic tasks. To verify this, we evaluate texture encoder $Enc^t$ on both lung disease classification and COVID-19 outcome prediction tasks.

\begin{table}[t]
\centering
\begin{tabular}{c|cccc}
\hline
Method & Params & 1\% labels ft & 10\% labels ft & linear evaluation \\
\toprule
MoCo-R18 & 11M & 61.1\% & 71.8\% & \textbf{73.0}\% \\
$Enc^t$ in LSAE & 5M & \textbf{61.3}\% & \textbf{73.2}\% & 69.0\% \\
\hline
\end{tabular}
\caption{\textbf{Comparison with MoCo (ResNet18) in self-supervised (linear evaluation) and semi-supervised settings}. Our method shows superior performance when compared with MoCo in semi-supervised settings while MoCo still ourperforms ours in the linear evaluation setting.}  \label{tab:moco}
\end{table}

\subsubsection{Disease Classification on ChestX-ray14.}
First, to test the extreme performance of our model, we finetune $Enc^t$ on the whole training set of ChestX-ray14 with all the available 14 disease labels, and report the class average AUC in Table~\ref{tab:cxr14}. We achieve competitive results compared with both fully-supervised pre-trained and self-supervised pre-trained methods (e.g., MoCo), albeit with a smaller model size. We further provide class-wise AUC in Table ~\ref{tab:cxr14-c}. 
% It is worth noting that our texture hypothesis may be weak for 4 of the 14 diseases, i.e., Pneumothorax, Cardiomegaly, Hernia, and Pleural Thickening. Structure variations are also accompanied with the progression of these 4 diseases. Interestingly, in Table ~\ref{tab:cxr14-c}, $Enc^t$ in LSAE indeed achieves inferior results in Cardiomegaly, Pleural Thickening, and Hernia, compared to ChestNet, which proves LSAE disentangles structure and texture effectively in turn.

%It should be noted that among the 14 diseases represented in the ChestX-ray14 dataset, we identify 4 that are more generally identified as changes in structure rather than texture on CXR: hernia, cardiomegaly, pleural thickening, and pneumothorax. 
It should be noted that, of the fourteen diseases classified in the ChestX-ray14 dataset, cardiomegaly and hernia are not lung pathologies. Indeed, as can be observed in Table ~\ref{tab:cxr14-c}, the texture encoder $Enc^t$ in LSAE achieves poorer results than ChestNet in these two classes. 
%This is in line with our hypothesis that LSAE disentangles structure and texture from CXRs, using texture to encode pathological information.  

Second, to compare with other state-of-the-art self-supervised and semi-supervised methods, we examine our model in three restricted settings, i.e., linear evaluation, 1\% labeled  data fine-tuning, and 10\% labeled data fine-tuning. Specifically, linear evaluation freezes the whole backbone network but only updates the final linear classifier. 1\% labeled data fine-tuning updates the whole network but with only 1\% of the labeled data. Similarly, 10\% labeled data fine-tuning updates the whole network with 10\% of the labeled data. We report results of the three settings of both our method and the recently-proposed method MoCo [\cite{he2020momentum}] in Table.~\ref{tab:moco}. As the results show, $Enc^t$ in LSAE can outperform MoCo in semi-supervised settings (i.e, 1\% labels finetuning and 10\% labels finetuning). However, MoCo still outperforms LSAE in the linear evaluation setting.

\begin{table}[t]
\centering
\begin{tabular}{cc}
\begin{tabular}{c|c|c|c}
\hline
BER(\%)$\downarrow$ & Inception v3 & $Enc^t$ in SAE & $Enc^t$ in LSAE \\
\hline
split 1 & 20.25 $\pm$ 1.46 & 20.25 $\pm$ 1.63 & \textbf{19.00} $\pm$ 1.84 \\
split 2 & 19.25 $\pm$ 3.67 & 20.50 $\pm$ 1.12 & \textbf{17.75} $\pm$ 1.66 \\
split 3 & 17.75 $\pm$ 3.48 & 14.00 $\pm$ 1.85 & \textbf{12.75} $\pm$ 2.15 \\
\hline
Avg & 19.08 & 18.25 & \textbf{16.50} \\
\hline
\end{tabular}
\\ \\
\begin{tabular}{c|c|c|c}
\hline
mAUC(\%)$\uparrow$ & Inception v3 & $Enc^t$ in SAE & $Enc^t$ in LSAE \\
\hline
split 1 & 85.45 $\pm$ 1.89 & 89.03 $\pm$ 2.16 & \textbf{89.17} $\pm$ 0.68 \\
split 2 & 86.02 $\pm$ 1.27 & 85.63 $\pm$ 1.77 & \textbf{87.07} $\pm$ 1.91 \\
split 3 & 89.12 $\pm$ 1.38 & 92.60 $\pm$ 1.25 & \textbf{95.00} $\pm$ 0.29 \\
\hline
Avg & 86.86 & 89.09 & \textbf{90.41} \\
\hline
\end{tabular}
\end{tabular}
\caption{\textbf{COVOC (Subset-1) Outcome Prediction Results}. We evaluate models with Balanced Error Rate (BER) and average AUC (mAUC). The texture encoder in LSAE surpasses Inception v3 by a large margin. It also outperforms the texture encoder in baseline model SAE, which demonstrates that better disentanglement does lead to better discrimination. We report mean and std of 5 random runs.} \label{tab:covoc}
\end{table}

\subsubsection{Outcome Prediction on COVOC.}
We further evaluate our model by transferring the pre-trained model to COVOC in two steps. First, on Subset-1 of COVOC, we evaluate the performance of $Enc^t$ in LSAE by adapting and finetuning it. Then, we test the Subset-1 model directly on Subset-2, which evaluates the generalizability of our model without any adaptation and finetuning.
\\
\noindent \textbf{Step 1: Adapting and Finetuning on Subset-1}. Considering the possible domain discrepancy between ChestX-ray14 and COVOC, we first adapt LSAE to the new domain by training to generate hybrid images on COVOC for 10K iterations. Then, we evaluate the texture encoder $Enc^t$ on the outcome prediction task by further finetuning. As COVOC is imbalanced, we report the 
ventilation prediction
Balanced Error Rate (BER)
together with mAUC in Table~\ref{tab:covoc}. Compared with Inception v3, $Enc^t$ reduces BER by 13.5$\%$, and improves mAUC by 4.1$\%$. When comparing with the texture encoder in baseline model SAE, LSAE also performs better. It demonstrates that better disentanglement leads to better discrimination.
\\
\noindent \textbf{Step 2: Directly Generalizing to Subset-2}. To further demonstrate the effectiveness of our method, we directly generalize the model trained on Subset-1 training set to Subset-2. As shown in Table~\ref{tab:cleve}, although the discrepancy between different domains decreases the performance relatively, LSAE can still achieve reasonable results and consistently outperform the baseline method.

\begin{table}[t]
\centering
\begin{tabular}{c|cc}
\toprule
& BER(\%)$\downarrow$ & mAUC(\%)$\uparrow$ \\
\toprule
Inception v3 (baseline) & 32.83 $\pm$ 7.34 & 76.01 $\pm$ 6.85 \\
\hline
$Enc^t$ in LSAE & 32.02 $\pm$ 4.22 & 77.58 $\pm$ 2.75 \\
$Enc^t$ in LSAE + Our Aug & \textbf{24.70} $\pm$ 2.77 & \textbf{79.94} $\pm$ 1.18 \\
\hline
\end{tabular}
\caption{\textbf{Generalization on COVOC Subset-2}. To better demonstrate the robustness of our method, we directly apply the model which is trained on Subset-1 training set to Subset-2 of COVOC.
Our method can achieve reasonable results without any adaptation. It also outperforms the baseline Inception v3.} \label{tab:cleve}
\end{table}

\subsection{Data Augmentation with Hybrid Images} \label{exp:fuse}
\begin{figure}[t]
\centering
\includegraphics[width=0.94\textwidth]{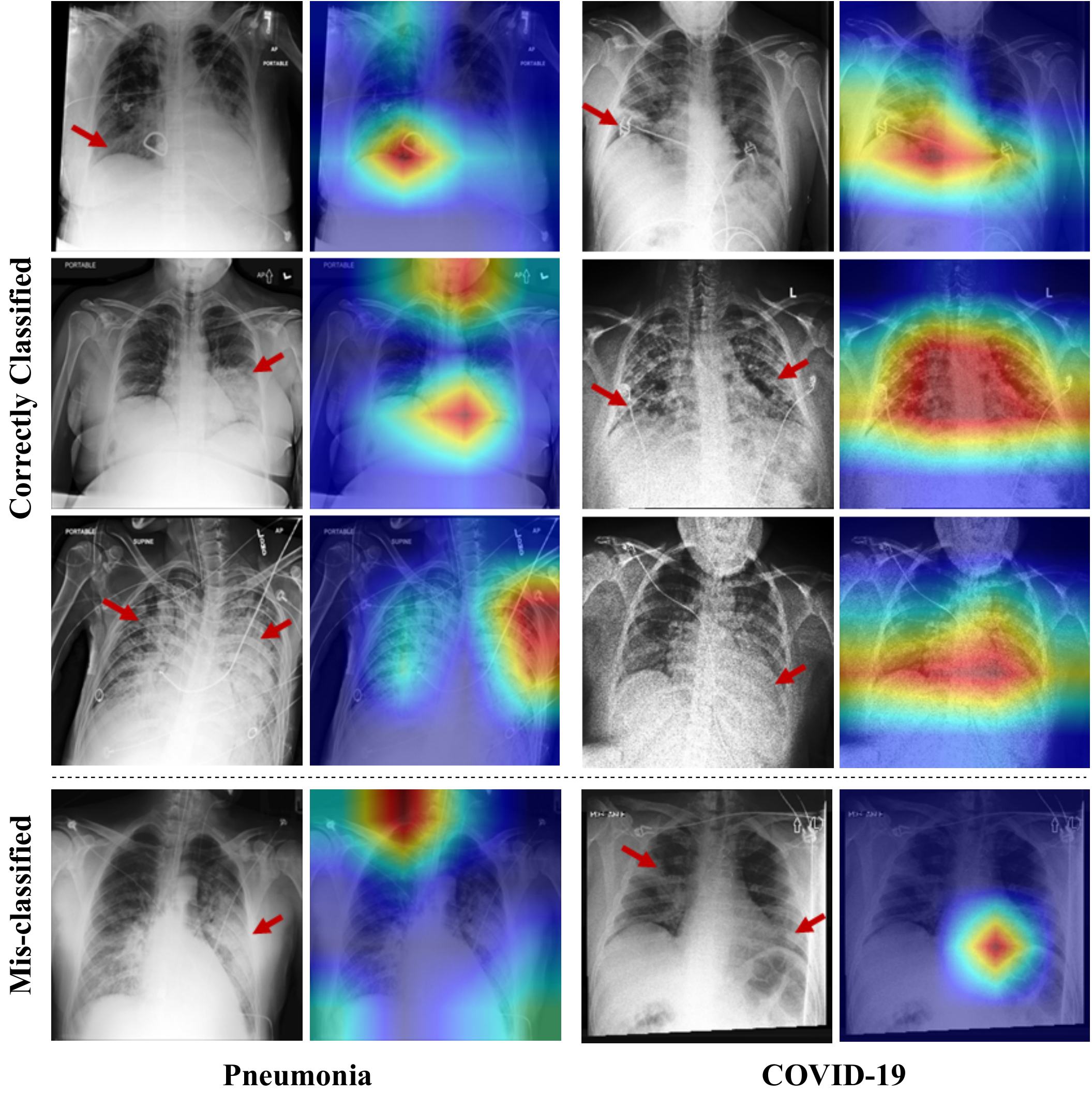}
\caption{\textbf{Grad-CAM results of $Enc^t$ in LSAE}. To get insights from the Grad-CAM visualizations, we collect corresponding annotations from radiologists (denoted by \textcolor{red}{red} arrows). We show the results of three correctly classified images in the first three rows, and one mis-classified image in the last row. Please refer to Sec.~\ref{vis} for detailed radiologists' analysis.} \label{fig:grad-cam}
\end{figure}
As described in Section~\ref{aug}, we generate hybrid images to augment the training data in COVOC. To control the training budget, we set $K=2$, i.e., the training data of COVOC is augmented to 2 times. To avoid introducing irrelevant diseases, we only sample structure images from the healthy lungs in ChestX-ray14. With the same experimental setup with Table~\ref{tab:covoc}, the augmentation method can reduce error rate further from 16.50$\%$ to 15.67$\%$, and improve mAUC from 90.41$\%$ to 92.04$\%$ on the ventilation prediction task on average over 3 splits. Moreover, to compare with others, we implement Mixup [\cite{zhang2017mixup}] independently for training on COVOC which achieves 16.41\% BER/90.82\% mAUC. Thus, our method still shows superior performance. As shown in Table~\ref{tab:cleve}, when directly generalizing on the Subset-2 data, our augmentation method can also help $Enc^t$ in LSAE get better results.

\subsection{Visualizations and Radiologist Interpretations} \label{vis}
\begin{figure}[t]
\centering
\includegraphics[width=\textwidth]{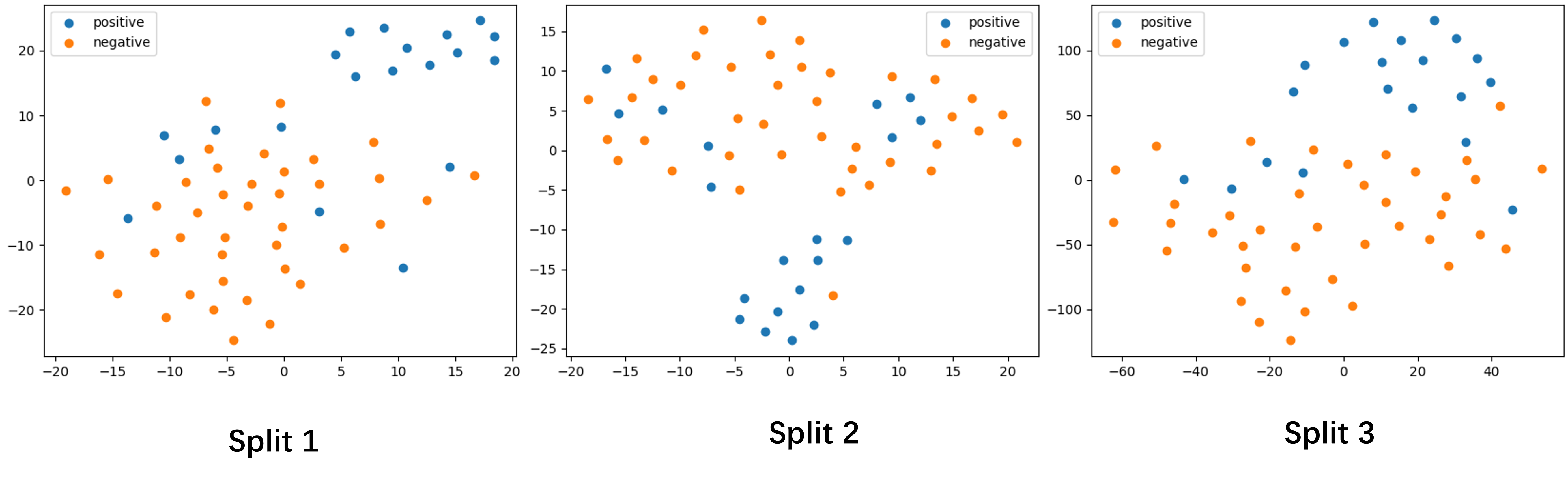}
\caption{\textbf{t-SNE results of features from $Enc^t$ in LSAE on COVOC (Subset-1)}. The three plots are feature clustering results of split1, split2, and split3 respectively. Most samples distribute in two clusters, one for \textcolor{blue}{positive} and the other for \textcolor{orange}{negative}, which indicates the effectiveness of the learned representations.} \label{fig:tsne}
\end{figure}
To better understand our model, we use Grad-CAM [\cite{selvaraju2017grad}] to highlight the key areas that contribute to the prediction on both ChestX-ray14 and COVOC. To get more insight, we also obtained radiologists' interpretations regarding the visualization results. Considering the time cost of annotations, in Chest-Xray14, we only pick up one COVID-related disease, i.e., Pneumonia. The Grad-CAM results are shown in the second and fourth columns in Fig~\ref{fig:grad-cam} while the corresponding radiologists' annotations are listed in the first and third columns. Generally speaking, for the cases which are correctly classified, Grad-CAM's highlighted regions match well with radiologists' annotations. In contrast, Grad-CAM results on misclassified images usually localize wrong areas of diseases. 
% Please refer to SM for individual analysis for each pair images from radiologists.

We further use t-SNE to visualize the representation vectors of $Enc^t$ in LSAE on the three splits of COVOC Subset-1. As shown in Fig~\ref{fig:tsne}, in the representation space, although not perfectly, most samples are distributed around two clusters, one for positive and the other for negative, which indicates the effectiveness of the learned representations.

\begin{figure}
\centering
\includegraphics[width=0.78\textwidth]{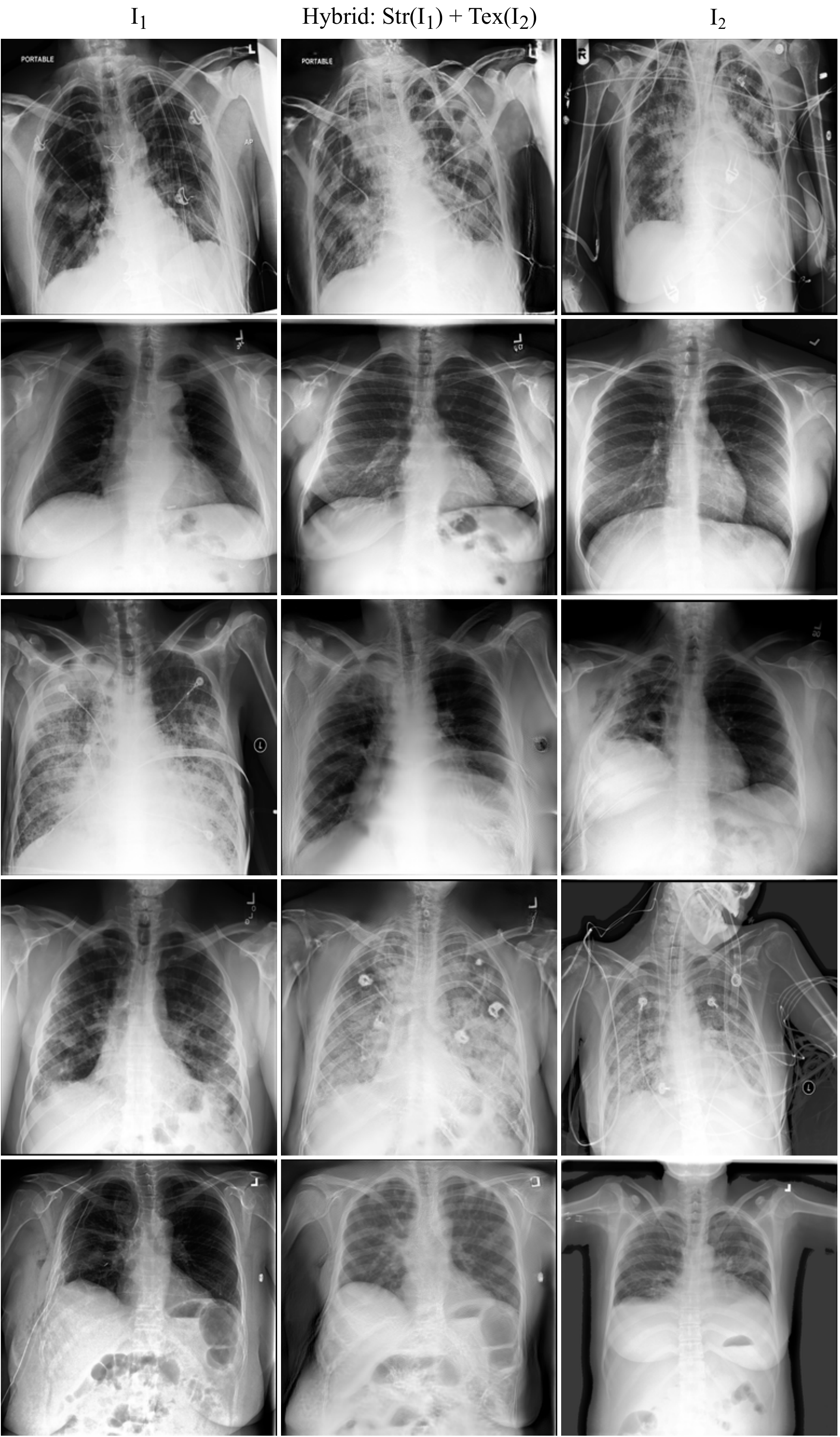}
\caption{\textbf{Lung swapping results on ChestX-ray14.} The first column shows $I_1$ images that provide structure templates. The third column shows $I_2$ images that provide texture targets. The middle column presents the hybrid images that mix the structure of $I_1$ and the texture of $I_2$.}
\label{more_results}
\end{figure}

\section{Discussion}
\subsection{Texture Hypothesis}
In this paper, we propose a self-supervised representation framework, Lung Swapping Autoencoder (LSAE), for CXRs. LSAE is based on the assumption that texture is tightly related to specific lung diseases while lung structures are regarded as less important in their diagnosis and prognosis. 
%Four lung diseases in ChestX-ray14, i.e., Pneumothorax, Cardiomegaly, Hernia, and Pleural thickening, can also cause lung structure changing, corroborated by our quantitative results. This indeed explains why the texture encoder $Enc^t$ in LSAE achieves inferior results in Table~\ref{tab:cxr14-c} for Cardiomegaly, Pleural Thickening, and Hernia. In turn, this proves LSAE can effectively disentangle the texture representation out of structure in a CXR. Generally speaking, our hypothesis still holds for most cases. 
The texture encoder $Enc^t$ in LSAE outperforms ChestNet in disease classification for all but three classes listed in the ChestX-ray14 dataset. Notably, two of these three disease classes (hernia and cardiomegaly) are not primarily pathologies of the lung, and therefore might not be expected to be suitable for analysis by LSAE.
%Notably, these three disease classes overlap with four classes of disease that we have identified as manifesting largely as structural rather than textural changes on CXR: hernia, cardiomegaly, pleural thickening, and pneumothorax. Hernias on CXR are identified as protrusions of gastrointestinal tissue in the retrocardial area rather than within lung fields~\cite{kohn2013guidelines}. Cardiomegaly refers to enlargement of the heart and would also not be significantly reflected in lung texture changes~\cite{doi:10.1080/16878507.2020.1756187}. Pleural thickening presents on CXR as thickening of the lung lining~\cite{saito2019pleural}. Finally, pneumothorax is the structural collapse of a lung within the chest cavity~\cite{kattea2015differentiating}. As demonstrated in Table ~\ref{tab:cxr14-c}, $Enc^t$ from LSAE underperforms ChestNet in cardiomegaly, pleural thickening, and hernia classification. The increased performance observed for pnuemothorax is not necessarily unexpected as the vacated chest cavity present due to a collapsed lung has substantially different textural information when compared with a normally inflated lung. 
% Consequently, the overall performance of $Enc^t$ in LSAE is better than many baselines with a much smaller model size on ChestX-ray14.

\subsection{Comparison with larger network} In Table~\ref{tab:cxr14} and Table~\ref{tab:moco}, we have compared with the state-of-the-art self-supervised learning method MoCo [\cite{he2020momentum}]. To guarantee a fair comparison, we re-implemented MoCo with a ResNet-18 (model size: 11M) backbone which has comparable model size as $Enc^t$ (model size: 5M). For discussion, we also re-implemented a MoCo-v2 with ResNet-50. Compared to MoCo-R18, MoCo-v2-R50 has a one time larger backbone network (model size: 23M), and introduces an auxiliary projection sub-network during training.
The performance of MoCo-v2-R50 on ChestX-ray14 across different supervision settings is listed below.
\begin{table}[h]
\centering
\begin{tabular}{c|cccccc}
\hline
Method & Params & epoch & 1\% & 10\% & 100\% & linear eval \\
\toprule
MoCo-v2-R50 & 23M & 200 & \textbf{65.1}\% & 72.8\% & \textbf{79.4}\% & \textbf{73.4}\% \\
$Enc^t$ in LSAE & 5M & 57 & 61.3\% & \textbf{73.2}\% & 79.0\% & 69.8\% \\
\hline
\end{tabular}
\caption{Results in self-supervised (linear evaluation) and semi-supervised settings} \label{tab:mocov2}
\end{table}
Although MoCo-v2-R50 shows best performance over almost all the supervision settings, it is worth mentioning that $Enc^t$ in LSAE can still win in the track of 10\% labeled data supervision with a 4 times smaller model size.

\subsection{Radiologist Survey} 1). In Table~\ref{fig:survey}, we list the survey results of each radiologist. Q1 and Q2 focus on the quality assessment of generated images. It is worth noting that radiologists are easier to be confused in the patch level than the image level. This is because there are more clues in the image level to provide evidence of synthesis, such as tags and pacemakers. When only showing patches, the radiologists can only depend on the in-lung texture to make a decision.

2). Along x-axis, we also provide radiologists' experiences, i.e., the residency years shown in the parentheses along x-axis. Except for Q1 where the most experienced radiologist also achieves the lowest confusion rate, we do not observe a significant correlation between the feedback and the experience in the other 3 questions.

Overall, the results of this radiologist survey support our proposal of Masked SFID as a measurement of texture similarity, and demonstrate the effectiveness of our image generation techinque.

\subsection{Hybrid Image Augmentation}
We have introduced a data augmentation method by generating hybrid images. There is a hyper-parameter $K$ in the method that indicates how many times the dataset is augmented. In this paper, enlarging the dataset by 2 times ($K = 2$) can reduce the error rate from 16.50\% to 15.67\%. However, when augmenting the dataset to a larger scale, we did not observe a consistent improvement of performance. We provide a possible explanation here for discussion: Our proposed augmentation strategy can increase the diversity of structure information. During training, high structural diversity can make the model more robust to the potential structure variations in test cases. However, it can hardly introduce more diversity of texture patterns. Therefore, the improvement can reach the limit after the model has exploited all the advantages of structural diversity.
Nonetheless, this hybrid image generation method and its promising application to COVID-19 datasets demonstrates its potential in future settings of rare or novel lung pathologies for which few images may exist for machine learning. 
\section{Conclusion}
To learn a transferable and clinically effective representation of CXRs in a self-supervised way, we proposed LSAE to disentangle texture from structure in CXR images, enabling analysis of disease-associated textural changes present in pulmonary diseases, such as COVID-19. Experiments on ChestX-ray14 show LSAE is able to learn an effective representation in fully-supervised, semi-supervised, and linear evaluation settings across many common pulmonary diseases. We also created a data augmentation technique, wherein synthetic images created using structural and textural information from two distinct CXRs can be used to expand datasets for machine learning applications, such as COVID-19 outcome prediction. Our resulting predictive model for mechanical ventilation in COVID-19 patients outperforms conventional methods and may have clinical significance in enabling improved decision making.

\section*{Acknowledgement}
Research reported in this publication was funded by the Office of the Vice President for Research and Institute for Engineering-Driven Medicine Seed Grants, 2019 at Stony Brook University.

\bibliography{elsarticle-template}

\end{document}